\documentclass[trackchanges, twocolumn]{aastex631}

\usepackage{mathtools}
\usepackage{nccmath}
\usepackage{graphicx}
\usepackage{float}
\usepackage{lipsum}
\usepackage{threeparttable} 
\usepackage{array}

\usepackage{amsmath}
\usepackage{booktabs}
\usepackage{multirow}

\newrobustcmd{\B}{\bfseries}

\renewcommand{\deg}{\ensuremath{^{\circ}}}

\shorttitle{BLERGs}
\shortauthors{Sankar et al.}
\graphicspath{{./}{figures/}}

\begin{document}
\title{When Jets Don’t Quench: Near-Infrared H$_{2}$ in Star Forming Low-Excitation Radio Galaxies}

\correspondingauthor{Swetha Sankar}
\email{ssanka10@jhu.edu}

\author[0000-0002-4419-8325]{Swetha Sankar}
\affiliation{Department of Physics and Astronomy, Bloomberg Center, Johns Hopkins University, 3400 N. Charles St., Baltimore, MD 21218, USA}

\author[0000-0003-4030-3455]{Andreea O. Petric}
\affiliation{Space Telescope Science Institute, 3700 San Martin Drive, Baltimore, MD 21218, USA}

\author[0000-0002-3007-0013]{Debora Pelliccia}
\affiliation{University of California Observatories, University of California, Santa Cruz, 1156 High Street, Santa Cruz, CA 95064, USA}

\author[0000-0001-6670-6370]{Timothy M. Heckman}
\affiliation{Department of Physics and Astronomy, Bloomberg Center, Johns Hopkins University, 3400 N. Charles St., Baltimore, MD 21218, USA}

\author[0000-0001-9122-9668]{Reiner M. J. Janssen}
\affiliation{Jet Propulsion Laboratory, California Institute of Technology, Pasadena, CA 91109, USA}

\author[0000-0002-3032-1783]{Mark Lacy}
\affiliation{National Radio Astronomy Observatory, Charlottesville, VA 22903, USA}

\author[0000-0001-5783-6544]{Nicole P. H. Nesvadba}
\affiliation{Université de la Côte d'Azur, Observatoire de la Côte d'Azur, CNRS, Laboratoire Lagrange, Bd de l'Observatoire, CS 34229, Nice cedex 4 F-06304, France}

\author[0000-0001-7883-8434]{Kate Rowlands}
\affiliation{Department of Physics and Astronomy, Bloomberg Center, Johns Hopkins University, 3400 N. Charles St., Baltimore, MD 21218, USA}
\affiliation{AURA for ESA, Space Telescope Science Institute, 3700 San Martin Drive, Baltimore, MD 21218, USA}

\author[0000-0002-9471-8499]{Pallavi Patil}
\affiliation{Department of Physics and Astronomy, Bloomberg Center, Johns Hopkins University, 3400 N. Charles St., Baltimore, MD 21218, USA}

\author[0000-0002-1428-7036]{Brian C. Lemaux}
\affiliation{Department of Physics and Astronomy, University of California, Davis, Davis, CA, USA}

\author[0009-0004-0844-0657]{Maya Skarbinski}
\affiliation{Department of Physics and Astronomy, Bloomberg Center, Johns Hopkins University, 3400 N. Charles St., Baltimore, MD 21218, USA}

\author[0009-0000-2546-1645]{Annie Giman}
\affiliation{Department of Physics and Astronomy, Bloomberg Center, Johns Hopkins University, 3400 N. Charles St., Baltimore, MD 21218, USA}

\author[0000-0003-3191-9039]{Justin A. Otter}
\affiliation{Department of Physics and Astronomy, Bloomberg Center, Johns Hopkins University, 3400 N. Charles St., Baltimore, MD 21218, USA}

\author[0000-0002-4261-2326]{Katherine Alatalo}
\affiliation{Department of Physics and Astronomy, Bloomberg Center, Johns Hopkins University, 3400 N. Charles St., Baltimore, MD 21218, USA}
\affiliation{Space Telescope Science Institute, 3700 San Martin Drive, Baltimore, MD 21218, USA}

\begin{abstract}
We present new Gemini/GNIRS near-infrared spectroscopic observations of eight low-redshift ($z<0.1$) blue low-excitation radio galaxies (BLERGs)---a rare subset ($\sim2.5\%$ of low-excitation radio galaxies; LERGs) that complicate the classical jet-mode AGN picture by combining radio activity with star-forming, gas-rich hosts. These star-forming BLERGs exhibit significant warm H$_2$ emission traced via ro-vibrational transitions at $T \sim 2,000$--$4,000$~K. We find that BLERGs span a broad range of mass-normalized warm H$_2$ luminosities ($L_{\rm H_2}/M_\star$), comparable to radio-emitting early-type galaxies, yet without a clear positive dependence on radio power. Instead, the strongest H$_2$ emission preferentially occurs in morphologically disturbed and advanced-merger systems, while compact radio sources ($\lesssim 20$~kpc) remain plausible sites of localized jet--ISM interaction. Together, these results suggest that merger-driven processes---including tidal shocks, gas inflows, and disturbed interstellar medium conditions---are the dominant drivers of warm molecular gas excitation in BLERGs, although localized jet-driven heating may contribute in individual systems. The compact radio morphologies, gas-rich hosts, and rarity of BLERGs are consistent with a short-lived evolutionary phase in which radio AGN activity coexists with an interaction-driven, molecular-rich interstellar medium prior to the onset of large-scale maintenance-mode feedback. Spatially resolved spectroscopy and higher-resolution radio imaging will be essential to disentangle the relative roles of mergers and jets in regulating the molecular gas of jet-mode AGN.

\end{abstract}


\section{Introduction} 
\label{sec:intro}

The tight connection between supermassive black holes (SMBHs) and their host galaxies---exemplified by the $M\textsubscript{\ensuremath{\bullet}}$--$\sigma$ relation \citep{ferr00, korm13}---has long motivated the view that feedback from active galactic nuclei (AGN) plays a central role in regulating galaxy growth. When accretion onto a SMBH releases energy into the surrounding interstellar or intracluster medium, it can heat or expel gas, suppressing star formation and preventing runaway and more moderate forms of cooling in massive halos \citep[e.g.,][]{mcna07,fabi12, Heckman2014}. This interplay between gas, star formation, and SMBH fueling underpins modern galaxy formation models and cosmological simulations, which require AGN feedback to reproduce key observables such as the high-mass end of the luminosity function, the quenching of massive galaxies, and the observed relations between the presence of AGN and warm molecular gas \citep{bowe06, crot06, Som2015, smol2017, petric2018, lamb2019, minsley2020}.

A central pillar of this feedback framework is the empirical division between accretion modes -- radiative-mode and jet-mode AGN \citep[e.g.,][]{hardc2007, smol2009, best12, smol2017}). Radiative-mode AGN--- observed as high-excitation radio galaxies (HERGs), radio-quiet QSOs, and Seyfert galaxies---are powered by radiatively efficient accretion from abundant cold gas, producing high-excitation emission lines and, in many cases, powerful accretion discs and dusty tori. In contrast, jet-mode AGN such as low-excitation radio galaxies (LERGs) are traditionally thought to be fueled by quasi-spherical, radiatively inefficient accretion flows from the hot gas halos of massive, red, passively evolving galaxies (e.g., \citealt{best12, Heckman2014}). These systems typically exhibit low Eddington-scaled accretion rates and launch low-power but long-lived jets that deposit mechanical energy into the surrounding medium, establishing a self-regulating feedback cycle that is especially effective in group and cluster environments.

However, emerging theoretical and observational results increasingly challenge this simple hot-mode picture for LERGs. Simulations of hot-halo accretion predict that thermal instabilities can precipitate cold filaments that ``rain'' onto the SMBH, potentially enabling LERGs to accrete cold gas under certain conditions \citep{Sha2012, gasp2013}. Observations likewise reveal a small but growing population of LERGs with detectable cold molecular gas reservoirs on kpc scales, underscoring the significance of multi-phase molecular interstellar medium studies of LERG hosts \citep[e.g.,][]{tremb2016, ruffa2019, petric2019}. The presence of such galaxy‑scale cold gas does not necessarily imply cold accretion onto the SMBH itself, since the immediate sub‑pc accretion flow may remain hot. Nevertheless, these findings imply that fueling in jet-mode AGN may not be restricted solely to hot gas and that the traditional HERG/LERG division may depend not only on the gas origin but also on the accretion modes (cold, radiatively efficient versus hot, radiatively inefficient) \citep{best12, hardc2018}, with substantial overlap between the two populations \citep{whit2018}.

Against this backdrop, rare examples of blue, gas-rich, star-forming LERGs present an especially intriguing challenge to canonical AGN feedback models, with their population density growing from the local Universe and peaking around cosmic noon \citep{konda2022}. In the low-redshift Universe, LERGs overwhelmingly inhabit massive, red-sequence hosts with old stellar populations; systems that deviate from this paradigm therefore provide key laboratories for probing non-standard fueling pathways and the interplay between jets and cold gas. Yet such ``blue LERGs'' (BLERGs) have received little systematic study, leaving a gap in our understanding of how jet-mode AGN operate in galaxies that possess substantial cold gas reservoirs.

Near-infrared spectroscopy of ro-vibrational H$_{2}$ transitions provides a sensitive probe of the warm molecular gas that coexists with cold gas reservoirs in BLERGs. These lines trace gas at temperatures of 1000–4000 K \citep[e.g.,][]{rose2013,sooj2004, mouri1994, dav2003, pado2022}, which can be excited by shocks, turbulence, or AGN-driven outflows \citep[e.g.,][]{nes2010, nesv06, ogle07, petric2018, minsley2020, lamb2019, UV2022, riff2020}, making them ideal for diagnosing the interplay between jets and the interstellar medium. Unlike optical emission lines, near-infrared H$_{2}$ lines are less affected by dust extinction, allowing us to detect molecular gas in regions of ongoing star formation. By combining near-IR H$_{2}$ measurements with CO observations (Janssen et al. in prep), we can probe multiple phases of the molecular ISM, shedding light on how jet-mode AGN operate in galaxies that retain substantial cold gas and active star formation.

In this work, we present new near-infrared spectroscopic observations of nine low-redshift ($z < 0.1$) blue low-excitation radio galaxies obtained with the GNIRS instrument on Gemini North. We begin by summarizing all known properties of our sample, the parent sample, and the reduction of the Gemini/GNIRS data in Section \ref{sec:obj}. Section \ref{sec:obs} details the emission line analysis of the near-infrared emission and H$_{2}$ lines. In Section \ref{sec:discussion}, we interpret these results in the context the role of mergers and environment and the influence of jets/ mechanical feedback to constrain the origin of the recent star formation. We summarize in Section \ref{sec:conclusions}. Throughout this paper we adopt the cosmological parameters H$_{0}$ = 69.6 km s$^{-1}$ Mpc$^{-1}$, $\Omega_{M}$ = 0.286, and $\Omega_{\Lambda}$ = 0.714 of the flat $\Lambda$CDM model \citep{bennett2014}.

\section{Sample Selection and Observation} 
\label{sec:obj}

\subsection{Parent Radio AGN Sample} \label{subsec:parentsamp}
The parent sample of radio-loud active galactic nuclei (RL AGN) used in this study is drawn from the catalog of \citet{best12}, which was constructed by cross-matching the Sloan Digital Sky Survey (SDSS) Data Release 7 (DR7; \citealt{abaz2009}) spectroscopic galaxy sample with the NRAO VLA Sky Survey (NVSS; \cite{cond98}) and the Faint Images of the Radio Sky at Twenty Centimeters survey (FIRST; \cite{beck95}). Optical host-galaxy properties are taken from the Max Planck Institute for Astrophysics--Johns Hopkins University value-added spectroscopic catalogs (MPA-JHU; \cite{brin2004}), which provide stellar masses, star formation rates, and spectral indices derived from SDSS spectroscopy \cite{kauf03a, kauf03b, kauf03c}. \citet{best12} identified a magnitude-limited sample of 18,286 radio sources with high completeness ($\approx95\%$) and reliability ($\approx99\%$). 

To ensure robust host-galaxy parameter estimates and minimize evolutionary effects, \citet{jans2012} further restricted this sample to galaxies with reliable redshifts in the range $0.03 \leq z \leq 0.3$ and stellar masses in the range $10^{10.25}\,M_\odot \leq M_* \leq 10^{12.0}\,M_\odot$. Point sources such as stars and quasars were removed through visual inspection, yielding a clean galaxy sample suitable for statistical analysis. Radio-loud AGN were separated from star-forming galaxies using a combination of optical emission-line diagnostics (the BPT diagram; \cite{bald81}), the relation between radio luminosity and H$\alpha$ luminosity, and the $L_{1.4\,\mathrm{GHz}}/M_*$ versus D$_\mathrm{n}4000$ plane (\cite{best05}). Host galaxies are classified as red, green, or blue using the strength of the 4000\,\AA\ break \citep[Dn4000;][]{balo99}: galaxies with Dn4000~\(\leq 1.45\) are classified as blue (114,947 systems), those with \(1.45 < \text{Dn4000} < 1.7\) as green (114,210 systems), and those with Dn4000~\(\geq 1.7\) as red (295,523 systems). After applying a minimum radio luminosity threshold of \(L_{1.4\,\text{GHz}} \geq 10^{23}\,\mathrm{W\,Hz}^{-1}\), \citet{jans2012} identify a total of 6,943 RL AGN, consisting of 6,691 LERGs — of which 171 are blue, star-forming systems — and 252  HERGs. Throughout this work, we adopt the RL AGN classifications, host-galaxy properties, and derived quantities as defined in \citet{jans2012} without re-deriving these parameters.

\subsection{Gemini/GNIRS BLERG Subsample} \label{subsec:oursamp}

\begin{table*}
\centering
\caption{Blue LERG Subsample Properties}
\label{tab:galaxy_params}
\begin{tabular}{lcccccc}
\hline
\hline
Name & RA (J2000) & Dec (J2000) & $z$ & $\log(M_\star/M_\odot)$ & $\log({\rm SFR}/M_\odot\,{\rm yr}^{-1})$ & 
$\log(L_{1.4\,{\rm GHz}}/{\rm erg\,s^{-1}\,Hz^{-1}})$ \\
\hline
J0754+3910 & 07:54:36.96 & +39:10:48.00 & 0.096 & 11.23 & 0.84 & 31.10 \\
J0028+0055 & 00:28:33.36 & +00:55:12.00 & 0.104 & 10.51 & 0.15 & 31.82 \\
J0809+3455 & 08:09:38.88 & +34:55:37.19 & 0.082 & 11.20 & 0.78 & 31.63 \\
J1040+2957 & 10:40:30.00 & +29:57:57.60 & 0.091 & 11.12 & 1.32 & 31.90 \\
J1056+1419 & 10:56:38.88 & +14:19:30.00 & 0.081 & 10.76 & 1.38 & 31.43 \\
J0056$-$0936 & 00:56:20.16 & $-$09:36:28.80 & 0.103 & 11.44 & 1.06 & 31.74 \\
J0829+1754 & 08:29:04.80 & +17:54:14.40 & 0.089 & 11.21 & 0.61 & 31.80 \\
J0758+2705 & 07:58:47.04 & +27:05:16.80 & 0.099 & 10.74 & 0.38 & 31.12 \\
\hline
\end{tabular}
\tablecomments{
Stellar masses and star formation rates are taken from the MPA--JHU DR7 catalog \citep{abaz2009, brin2004}. 
Radio luminosities are rest-frame 1.4 GHz monochromatic luminosities
computed from integrated NVSS flux densities \citep{cond98} assuming a spectral index 
$\alpha = 0.7$.}
\end{table*}

We focus exclusively on the 171 blue LERGs---hereafter blue low-excitation radio galaxies (BLERGs)---identified within the \citet{jans2012} LERG population. With Dn4000~\(\leq 1.45\), these systems host young stellar populations and ongoing star formation, placing them in stark contrast to the quiescent red LERG majority. To enable high signal-to-noise near-infrared spectroscopy and maximize spatial resolution, we restrict our targets to the low-redshift (\(z < 0.1\)) tail of this population. We further prioritize BLERGs with existing Eight MIxer Receiver (EMIR) IRAM30m \citep{carter2012} CO observations, enabling a multi-phase characterization of the molecular ISM through direct comparison between warm molecular gas traced by near-infrared H$_2$ ro-vibrational emission and cold molecular gas traced by CO in Janssen et al. in prep. Ten BLERGs satisfying these criteria were initially selected for Gemini/GNIRS observations. Two targets were subsequently excluded due to insufficient data quality yielding a final sample of eight BLERGs whose properties are listed in Table \ref{tab:galaxy_params}. These galaxies span stellar masses of \(10^{10.5}\,M_\odot \leq M_* \leq 10^{11.8}\,M_\odot\) and radio luminosities of \(24.1~\mathrm{W\,Hz}^{-1} \leq \log L_{1.4\,\text{GHz}} \leq 24.9~\mathrm{W\,Hz}^{-1}\).

\subsection{Gemini/GNIRS Observation $\&$ Reduction}
We observed eight blue low-excitation radio galaxies (BLERGs) with the Gemini Near-Infrared Spectrograph \citep[GNIRS;][]{elias2006} on the Gemini North 8.1~m telescope in queue mode during the 2019B semester (program ID: GN-2019B-Q-321). The observations were designed to detect and characterize ro-vibrational H$_2$ emission tracing warm molecular gas potentially excited by shocks associated with radio jets. All targets were observed using the cross-dispersed (XD) mode, which provides simultaneous wavelength coverage from approximately 0.8--2.5~$\mu$m, enabling the detection of multiple H$_2$ transitions in the $K$ band as well as diagnostic lines such as [Fe\,\textsc{ii}] and hydrogen recombination lines. We employed the 32~l~mm$^{-1}$ grating with the short blue camera (0.15\arcsec/pix), and 0.3\arcsec wide slit, yielding a spectral resolving power of $R \simeq$~1800 (corresponding to a velocity resolution of $\simeq$~170~km~s$^{-1}$). This resolution is sufficient to resolve both narrow and moderately broadened H$_2$ line components and to compare the kinematics of the warm molecular gas with existing CO measurements.

Each galaxy was observed for a total on-source integration time of 1800~s, divided into individual exposures and nodded along the slit in an ABBA pattern to enable accurate sky subtraction. Because the BLERGs are spatially extended in the near-infrared, sky subtraction was performed using object nodding rather than separate sky frames. Including acquisition, telluric standard observations, and overheads, the total time per source was approximately 1~hr. Observations were carried out under moderate conditions (cloud cover $\leq 70\%$, image quality $\leq$~0.85$''$), consistent with Gemini Band~3 requirements. 
Telluric standard stars of spectral type A0V were observed close in time and airmass to each science target, using the same instrumental setup and dither pattern, to enable correction for atmospheric absorption and flux calibration. 

The GNIRS data were reduced using the \texttt{PypeIt} spectroscopic pipeline \citep{joss20, P24zenodo} together with custom Python routines for telluric correction and absolute flux calibration. Raw frames were corrected for detector artifacts and flat-fielded, after which \texttt{PypeIt} was used for order tracing, sky subtraction, and wavelength calibration. Wavelength solutions were derived from atmospheric OH sky emission lines in the science frames, which provide a more robust calibration for GNIRS data than arc lamps due to instrument flexure.

The extracted echelle orders were trimmed by 10$\%$ at each end to remove low-sensitivity regions near order boundaries. Reference spectra were generated using a 9500~K blackbody, with intrinsic hydrogen absorption features masked. The telluric response function, defined as the ratio of the observed standard-star spectrum to the model, was smoothed and interpolated across masked regions before being applied to the science spectra on an order-by-order basis. Flux calibration was performed by scaling the telluric-corrected spectra to match 2MASS $J$, $H$, and $K_s$ photometry. Synthetic photometry was computed by integrating the spectra through the corresponding filter transmission curves, using only orders with significant wavelength overlap. A single global scaling factor was derived for each galaxy from a weighted mean across bands and applied uniformly to all orders, with uncertainties propagated from both the photometry and synthetic flux measurements. The final spectra span $0.8$–$2.5~\mu$m and achieve typical signal-to-noise ratios of $S/N \gtrsim 5$ in the strongest H$_2$ ro-vibrational lines.

\subsection{Emission Line Fitting}
Emission-line measurements were carried out on the flux-calibrated GNIRS spectra using a custom Python fitting routine. Per galaxy, the continuum was modelled with a polynomial of degree two to four---selected based on the signal-to-noise and spectral curvature of each target---using iterative sigma-clipping weighted least-squares fitting with explicit masking of expected line regions, and subtracted prior to line analysis. We then searched for near-infrared emission lines including ro-vibrational H$_2$ transitions from the 1--0 and 2--1 vibrational bands, hydrogen recombination lines from the Brackett series (Br$\delta$ at 19451\,\AA, Br$\gamma$ at 21661\,\AA), He\,I, [Fe\,II], and coronal lines such as [Si\,{\sc vi}] $\lambda$19641\,\AA\, restricting the analysis to transitions whose redshifted wavelengths fall within the observed spectral range of 17000--24150\,\AA. Emission line candidates were identified by searching the continuum-subtracted spectrum for peaks exceeding a signal-to-noise threshold of $1.8$--$3.0\,\sigma$ (adjusted per target based on data quality), using a noise estimate derived from the median absolute deviation of the continuum-subtracted flux.

Individual lines were initially fit independently with single Gaussian profiles over windows of $\pm 50\,\mathrm{\AA}$ around each line center, yielding line fluxes, centroids, and velocity dispersions as free parameters. Where multiple H$_2$ ro-vibrational transitions were detected within the same vibrational band, line parameters were line tied, a procedure in which all lines within a given band are constrained to share a common line-of-sight velocity and velocity dispersion, with only their amplitudes left as free parameters. This reduces the degrees of freedom, improves fit stability in lower signal-to-noise regions, and ensures physically self-consistent kinematics across transitions arising from the same emitting gas. This condition was typically met for the 1–0 band in our sources, while higher-vibrational transitions were detected in isolation and were therefore fit independently. For J1056+1419, the [Si\,{\sc vi}] $\lambda$19641\,\AA\ profile exhibited significant excess flux in the line wings relative to a single Gaussian component. We therefore fit this line with multiple Gaussian components with a shared kinematic center but a free velocity dispersion.

Flux uncertainties were estimated empirically from the root-mean-square noise in line-free regions immediately adjacent to each line, scaled by the effective line width in pixels, which provides a more conservative and robust error estimate than the formal covariance matrix of the fit. Velocity widths were converted to rest-frame values assuming Gaussian profiles. Lines with amplitudes exceeding their formal uncertainties by more than $3\sigma$ were classified as secure detections; lower-significance features ($2$--$3\sigma$) were treated as marginal detections and flagged accordingly. For lines not meeting the $2\sigma$ threshold, $3\sigma$ upper limits on the integrated flux were computed as $3 \times \mathrm{RMS} \times \sqrt{2\pi}\,\sigma_\lambda$ is the line width in wavelength units taken from the most dynamically relaxed system in our sample, J0056--0936, with a secure H2(1–0)S(1) detection as a representative upper limit on the expected line width.

\begin{deluxetable*}{lcccccccccccccccc}
\rotate
\tablecaption{Near-IR Emission Line Fluxes and Widths\label{tab:linefluxes}}
\tablewidth{0pt}
\tablehead{
\colhead{Line} &
\multicolumn{2}{c}{J0754+3910} &
\multicolumn{2}{c}{J0028+0055} &
\multicolumn{2}{c}{J0809+3455} &
\multicolumn{2}{c}{J1040+2957} &
\multicolumn{2}{c}{J1056+1419} &
\multicolumn{2}{c}{J0056$-$0936} &
\multicolumn{2}{c}{J0829+1754} &
\multicolumn{2}{c}{J0758+2705} \\
\colhead{} &
\colhead{Flux} & \colhead{FWHM} &
\colhead{Flux} & \colhead{FWHM} &
\colhead{Flux} & \colhead{FWHM} &
\colhead{Flux} & \colhead{FWHM} &
\colhead{Flux} & \colhead{FWHM} &
\colhead{Flux} & \colhead{FWHM} &
\colhead{Flux} & \colhead{FWHM} &
\colhead{Flux} & \colhead{FWHM}
}
\startdata
Br$\delta$ & $<3.93$ & 445 & $<8.60$ & 445 & $<4.19$ & 445 & $9.15\pm0.77$ & 653 & $10.3\pm1.8$ & 222 & $<1.15$ & 445 & $<3.42$ & 445 & $<3.00$ & 445 \\
H$_2$(1--0)S(3) & $<3.99$ & 445 & $<5.76$ & 445 & $<2.43$ & 445 & $18.9\pm0.7$ & 662 & $277\pm21$ & 444 & $<0.80$ & 445 & $<3.79$ & 445 & $<3.04$ & 445 \\
H$_2$(1--0)S(2) & $<2.47$ & 445 & $<7.56$ & 445 & $<3.92$ & 445 & $6.65\pm0.29$ & 662 & $47.3\pm1.1$ & 444 & $<1.74$ & 445 & $<3.52$ & 445 & $<3.84$ & 445 \\
H$_2$(1--0)S(1) & $<3.99$ & 445 & $<11.3$ & 445 & $<5.02$ & 445 & $9.90\pm0.67$ & 662 & $103\pm3$ & 444 & $3.79\pm0.40$ & 445 & $<4.18$ & 445 & $<3.48$ & 445 \\
H$_2$(2--1)S(1) & $<5.18$ & 445 & $<12.4$ & 445 & $18.2\pm2.0$ & 602 & $<1.72$ & 445 & $9.73\pm2.05$ & 512 & $<1.50$ & 445 & $2.35\pm0.56$ & 182 & $<3.95$ & 445 \\
Br$\gamma$ & $<4.19$ & 445 & $<15.4$ & 445 & $5.08\pm1.28$ & 187 & $<2.44$ & 445 & $40.8\pm2.5$ & 603 & $<2.03$ & 445 & $5.10\pm0.44$ & 596 & $<3.93$ & 445 \\
\enddata
\tablecomments{
Fluxes are reported in units of $10^{-16}$ erg s$^{-1}$ cm$^{-2}$.
FWHM values are in km s$^{-1}$. Upper limits are 3$\sigma$ and have been
rescaled to a common assumed FWHM of 445 km s$^{-1}$, corresponding to
the measured H$_2$(1--0)S(1) width in J0056$-$0936. For detections, the
measured FWHM is listed. Where Gaussian $\sigma$ values were reported,
FWHM values were computed using ${\rm FWHM}=2.355\sigma$.
}
\end{deluxetable*}

\begin{deluxetable*}{lcccccc}
\tablecaption{[Fe\,{\sc ii}] and [Si\,{\sc vi}] Line Measurements\label{tab:sivi_feii}}
\tablewidth{0pt}
\tablehead{
\colhead{Name} &
\colhead{[Fe\,{\sc ii}] Flux} &
\colhead{[Fe\,{\sc ii}] FWHM} &
\colhead{[Si\,{\sc vi}] Flux} &
\colhead{[Si\,{\sc vi}] FWHM} &
\colhead{[Si\,{\sc vi}] Wing Flux} \\
\colhead{} & \colhead{} & \colhead{(km s$^{-1}$)} & \colhead{} & \colhead{(km s$^{-1}$)} & \colhead{} 
}
\startdata
J1056+1419 & $111\pm1.67$ & 715 & $150\pm8.52$ & 664 & $47.3$ \\
J1040+2957 & $<3.70$ & 235 & $4.45\pm0.89$ & 657 & \nodata  \\
\enddata
\tablecomments{
Fluxes are in units of $10^{-16}$ erg s$^{-1}$ cm$^{-2}$. Upper limits are 3$\sigma$; for the [Fe,{\sc ii}] non-detection, the upper limit assumes an unresolved line width of 235 km s$^{-1}$, corresponding to the instrumental resolution. FWHM values are in km s$^{-1}$. For J1056+1419, the listed [Si,{\sc vi}] flux and FWHM correspond to the central gaussian component. The additional [Si,{\sc vi}] flux is the summed contribution of the red- and blue-shifted wing components identified in the multi-component fit. Gaussian $\sigma$ values were converted using ${\rm FWHM}=2.355\sigma$. }
\end{deluxetable*}

\section{Analysis}
\label{sec:obs}

\subsection{Emission Line Measurements $\&$ Detections}
\label{subsec:emline}
\begin{figure*}
    \centering
    \includegraphics[width=1.0\textwidth]{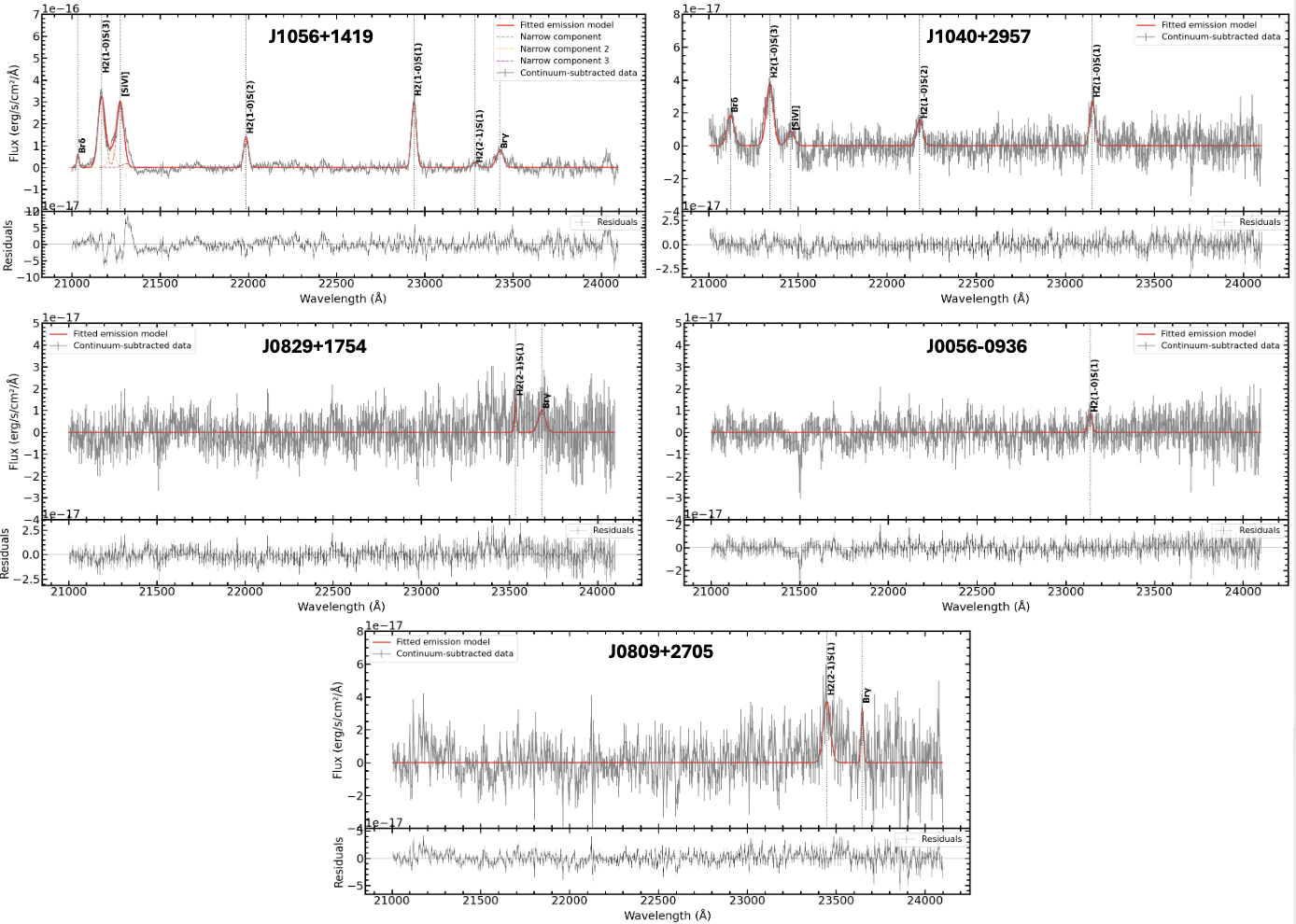}\\
    \caption{Continuum-subtracted observed K-band GNIRS spectra and emission line fits for the five targets in our sample with detected emission features. In each panel, the grey curve shows the continuum-subtracted data with error bars, and the red curve shows the total fitted emission line model. Vertical dotted lines mark the expected positions of detected transitions at the systemic redshift of each target. J1056+1419 (top left, $z = 0.08$), the most line-rich system in the sample, showing secure detections of several lines. The [SiVI] profile is fit with a multi-component redshifted (purple) and blueshifted (orange) Gaussian model sharing a common kinematic centre. The lower panel beneath each spectrum shows the fit residuals. The four targets in the sample with no detected emission features above the noise threshold are not shown; stacking of these spectra yields no significant signal.
}
    \label{fig:gnirsH2}
\end{figure*}

The continuum-subtracted spectra and best-fit emission line models for all five targets with detections are shown in Figure~\ref{fig:gnirsH2}. Across the full sample of eight targets, five systems yield near-infrared emission line detections, of which three show secure ro-vibrational H$_2$(1--0) emission; the remaining four show no detections above the noise threshold, and stacking of the non-detected systems yields no significant signal. Of the five detected systems, the sample spans a wide range in line richness and excitation complexity. Measured emission-line fluxes and widths for each galaxy are presented in Table \ref{tab:linefluxes}.

\subsubsection{J1056+1419}
\label{subsubsec:J1056+1419}
J1056+1419 is the most line-rich target in the sample, with secure detections of Br$\delta$, H$_2$(1--0)S(3), H$_2$(1--0)S(2), H$_2$(1--0)S(1), H$_2$(2--1)S(1), [Fe\,{\sc ii}]~1.644~$\mu$m, and [Si\,{\sc vi}]~1.9641~$\mu$m. All detected H$_2$ 1--0 transitions were fit with kinematically tied profiles anchored to H$_2$(1--0)S(1), which yields a systemic redshift of $z = 0.0811$ and a velocity dispersion of $\sigma = 180\,\mathrm{km\,s}^{-1}$.

Residual broad emission is detected in the wings of [Si\,{\sc vi}], well-described by two Gaussian components with $\sigma \approx 300\,\mathrm{km\,s}^{-1}$ centered at $21227.9\,\mathrm{\AA}$ and $21304.6\,\mathrm{\AA}$, respectively, with a combined flux of $4.73 \times 10^{-15}~\mathrm{erg\,s}^{-1}\,\mathrm{cm}^{-2}\,\mathrm{\AA}^{-1}$. The physical origin of this broad emission is not uniquely constrained---both a hot turbulent gas component and a kinematically disturbed outflow are consistent with the observed line widths. The simultaneous detection of H$_2$(2--1)S(1) provides an important additional excitation diagnostic. The observed H$_2$(2--1)S(1)/H$_2$(1--0)S(1) flux ratio of $0.094$ is consistent with pure thermal excitation, for which a ratio of $\sim 0.1$--$0.2$ is expected, while values of $\sim 0.5$--$0.6$ indicate UV fluorescent pumping by OB stars in photodissociation regions \citep{mouri1994, reun2002}. This is consistent with collisional excitation of warm molecular gas at $T \sim 2000\,\mathrm{K}$, most plausibly heated by shocks. 

We detect both [Fe\,{\sc ii}] at $1.644\,\mu\mathrm{m}$ and [Si\,{\sc vi}] at $1.9641\,\mu\mathrm{m}$ (Table \ref{tab:sivi_feii}), tracers of distinct physical mechanisms that together paint a coherent picture of coexisting AGN and shock activity. The [Fe\,{\sc ii}] $1.644\,\mu\mathrm{m}$ line further suggests the presence of shocks, as iron is strongly depleted in the gas-phase ISM and shocks are able to destroy Fe-rich dust grains, producing an excess of [Fe\,{\sc ii}] relative to photodissociation region models \citep{mour00}. [Si\,{\sc vi}] has an ionization potential of $167\,\mathrm{eV}$ and cannot be produced by stellar radiation fields, making it a known tracer of AGN photoionization or fast shocks \citep{lamp2017, rodri2004, rodr11}. The H$_2$(1--0)S(1)/[Si\,{\sc vi}] and [Si\,{\sc vi}]/Br$\gamma$ flux ratios of $0.69$ and $3.65$, respectively, are comparable to values observed in Mrk\,273, the host of a powerful AGN \citep{Viv2013}. In Mrk\,273, spatially resolved adaptive-optics-assisted NIR spectroscopy indicates that AGN photoionization---or alternatively fast shocks exceeding $\sim 550\,\mathrm{km\,s}^{-1}$---is the most likely excitation mechanism for [Si\,{\sc vi}] \citep{Viv2013}. 

In our spectrum, the [Si~VI] emission is well detected and is best described by a multi-component Gaussian model consisting of a narrow core plus additional lower-flux components. We emphasize, however, that the observed [Si~VI] profile shows broader low-level wings than are fully reproduced by the adopted multi-component Gaussian model, so the quoted uncertainty should be interpreted as the uncertainty on the adopted fit rather than a complete measure of the line-shape complexity. The adopted fit yields a total integrated [Si~VI] flux of $(1.97 \pm 0.11) \times 10^{-14}$ erg s$^{-1}$ cm$^{-2}$, with the narrow component contributing $1.49 \times 10^{-14}$ erg s$^{-1}$ cm$^{-2}$ and the additional components together contributing $\sim 24\%$ of the total flux. The core is centered at $21274.6$~\AA\ with $\sigma \approx 282$ km s$^{-1}$ (FWHM $\approx 664$ km s$^{-1}$). The characteristic velocity widths of the fitted [Si\,{\sc vi}] components reach values exceeding the $\gtrsim 550~{\rm km\,s^{-1}}$ shock velocities required to collisionally excite [Si\,{\sc vi}], making shock excitation a plausible contributor.



\subsubsection{J1040+2957}
J1040+2957 shows secure detections of Br$\delta$, H$_2$(1--0)S(3), H$_2$(1--0)S(2), H$_2$(1--0)S(1), and [Si\,{\sc vi}]~1.9641~$\mu$m (see Table \ref{tab:sivi_feii} for [Si\,{\sc vi}] flux and width). The lower-$J$ transitions S(1) and S(2) display clean, single-component profiles, excluding a significant secondary kinematic component in the molecular gas. A distinct narrow emission feature at $\lambda_{\rm obs} \sim 21620$~\AA{} is detected immediately redward of H$_2$(1--0)S(3), well separated from the S(3) peak. This is identified as [Si\,{\sc vi}] at the systemic redshift. 

Unlike J1056+1419, no multi-component structure is observed in [Si\,{\sc vi}], indicating a less disturbed gas reservoir. The presence of [Si\,{\sc vi}] in this target is once again associated with AGN or energetic shock excitation (see Section~\ref{subsubsec:J1056+1419}).

\subsubsection{J0056$-$0936}
\label{subsubsec:J0056}
J0056$-$0936 shows a marginal detection of H$_2$(1--0)S(1) only. The non-detection of higher-$J$ ro-vibrational transitions in the unaffected spectral regions is consistent with thermally excited molecular gas at moderate temperatures of $T \sim 1000$--$2000$~K, where the S(1) line carries the dominant flux and higher-$J$ lines fall below the detection threshold \citep{Pak2004}.

\subsubsection{J0829+1754 and J0809+2705}
J0829$+$1754 and J0809$+$2705 show no ro-vibrational H$_2$(1--0) transitions but instead yield secure detections of H$_2$(2--1)S(1) and Br$\gamma$ in both systems. The detection of H$_2$(2--1)S(1) without any accompanying H$_2$(1--0) band emission places these systems in a physically distinct regime from the remainder of the sample. The non-detection of the 1--0 band transitions places lower limits on the H$_2$(2--1)S(1)/H$_2$(1--0)S(1) flux ratio of $>0.23$ and $>0.79$ in J0829$+$1754 and J0809$+$2705, respectively, where the upper limit on H$_2$(1--0)S(1) in each system was computed using the local RMS noise and a representative line width of $\sigma = 280~\text{km}~\text{s}^{-1}$, consistent with H$_2$(1--0)S(1) detections across the remainder of the sample. For comparison, pure thermal excitation predicts H$_2$(2--1)S(1)/H$_2$(1--0)S(1) $\sim 0.1$--$0.2$, while UV fluorescent pumping in PDRs yields ratios of $\sim 0.5$--$0.6$ \citep{mouri1994}. 

The lower limit for J0829$+$1754 ($>0.23$) marginally exceeds the thermal range, while that for J0809$+$2705 ($>0.79$) substantially exceeds both thermal and fluorescent predictions. However, we caution that the expected H$_2$(1--0)S(1) wavelength of $\sim 22915~\text{\AA}$ at $z = 0.08$ coincides with a region of potentially compromised telluric transmission, which could artificially suppress any underlying emission and inflate the inferred limit for J0809$+$2705. The presence of Br$\gamma$ emission in both systems confirms the presence of warm ionized gas. Deeper observations targeting the H$_2$(1--0) band are required to determine whether the 2--1 detections reflect genuinely unusual excitation conditions---such as UV fluorescent pumping or X-ray heating---or whether the 1--0 transitions are simply below the noise floor of the current data.

\subsection{H$_2$ Excitation Diagrams}
\label{subsec:h2excite}

\begin{figure}
    \centering
    \includegraphics[width=0.45\textwidth]{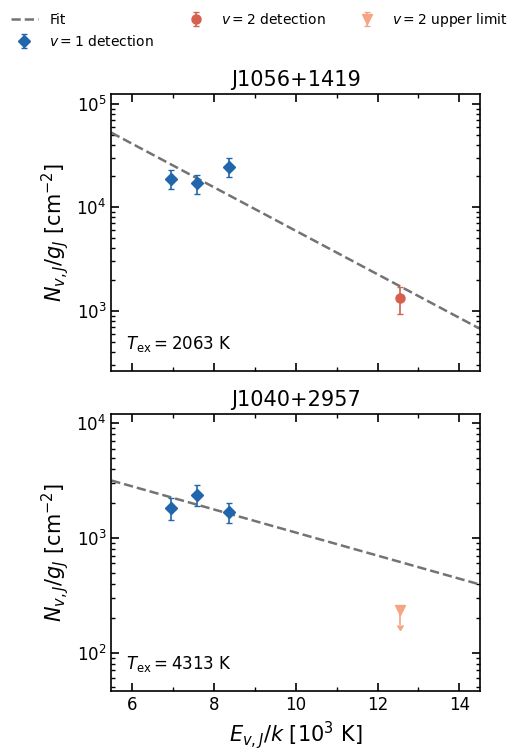}\\
     \caption{H$_2$ excitation diagrams for J1056+1419 \textit{(top)} and J1040+2957 \textit{(bottom)}. Blue diamonds are detections of $v=1$ rotational–vibrational lines, plotted as $\log\!\left(N_{v,j}/g_{v,j}\right)$ in cm$^{-2}$ (not normalized) against upper-level energy $E_{v,j}/k$ in units of $10^{3}\ \mathrm{K}$. The red circle in the top panel is a $v=2$ detection; the salmon downward triangle shows $v=2$ upper limits. Red vertical bars indicate uncertainties. The grey dashed line in each panel is the best-fit single-temperature Boltzmann distribution; the fitted excitation temperature $T_{\mathrm{ex}}$ is listed in the lower-left of each panel.}
     \label{fig:Tex}
    
\end{figure}

Using the H$_2$ line fluxes $f$ (W\,m$^{-2}$) listed in Table~\ref{tab:linefluxes}, 
we calculated the corresponding molecular column densities in the upper 
ro-vibrational $(v,\,j)$ levels of the detected transitions, given by

\begin{equation}
    N_{v,j} \approx \frac{4\pi\,f}{A_{v,j}\,\Omega\,h\nu_{v,j}},
\end{equation}

\noindent where $A_{v,j}$ is the radiative decay rate, $h\nu_{v,j}$ is 
the photon energy of the transition, and $\Omega = 0.13$\,arcsec$^2$ is the 
solid angle of the extraction aperture. We assume that the quadrupole 
transitions are optically thin and adopt the radiative $A$ coefficients 
from \citet{Wolniewicz1998}. Figure~\ref{fig:Tex} presents plots of 
$\log\left(N_{v,j}/g_{v,j}\right)$ --- with $g_{v,j}$ the statistical 
weight of the $(v,\,j)$ level --- against the upper level energy $E_{v,j}$ 
(in units of $10^3$\,K) for the two targets with sufficient H$_2$ line 
detections to construct excitation diagrams: J1056+1419 and J1040+2957. 
The population distributions are normalised to $N_{1,3}/g_{1,3}$, as 
derived from the H$_2$(1--0)\,S(1) line. For gas in local thermodynamic 
equilibrium at a single kinetic temperature, the level populations follow 
a Boltzmann distribution and the points in such a diagram are expected to 
fall along a straight line of slope $-1/T_\mathrm{ex}$ 
\citep[e.g.][]{Reunanen2002}.

The best-fit single-temperature models yield excitation temperatures of $T_{\rm ex} = 2063$~K for J1056+1419 and $T_{\rm ex} = 4313$~K for J1040+2957. These values are at or significantly higher than the $T \sim 1000$--$2000$~K characteristic of slow C-type shocks and quiescently heated molecular gas \cite{shul82, Hollenbach1989}, and are instead more consistent with fast J-type or magnetically mediated shocks with velocities $v_s \gtrsim 20$--$40$~km~s$^{-1}$, which can heat molecular gas to temperatures of several thousand Kelvin \cite{Hollenbach1989}. We caution that a single-temperature fit to a multi-component gas distribution will systematically overestimate $T_{\rm ex}$; the true gas temperature distribution likely spans a range of values, with the fitted $T_{\rm ex}$ representing an emission-weighted upper envelope. 


In J1056+1419, the excitation diagram shows a systematic departure from the single-temperature fit: the highest-energy $v = 1$ points lie above the best-fit line, indicating excess population at higher energies relative to a pure single-temperature Boltzmann distribution. Additionally, the $v = 2$ detection at high upper-level energy ($E_{v,J}/k \approx 12{,}500$~K) falls below the extrapolated fit, broadly consistent with the overall slope but providing an important anchor at high excitation. This pattern is the characteristic signature of a multi-temperature molecular gas structure, in which a cool shocked component ($T \sim 1000$--$2000$~K) dominates the lower-$J$ transitions while a hotter shock-heated component contributes disproportionately to the higher-$J$ lines (e.g., \cite{reun2002, davi03, mazz13}). Such multi-temperature structures are commonly observed in interacting and merging systems, where tidal forces drive shocks of varying velocities into the molecular gas, naturally producing a superposition of thermal components \cite{guil09, ogle10, petric2018}.

For J1040+2957, the $v = 1$ excitation diagram points are broadly consistent with the single-temperature fit within the measurement uncertainties, with the derived excitation temperature once again high, indicative of shock-heated gas. At high upper-level energy ($E_{v,J}/k \approx 12{,}500$~K), only an upper limit is available for the $v = 2$ transition; this upper limit lies below the extrapolated single-temperature fit, and suggests a multi-temperature structure. We note that the narrow range of upper-level energies sampled among the $v = 1$ detections in J1040+2957 ($E_{v,J}/k \sim 6500$--$9500$~K) means the slope of the excitation diagram — and therefore $T_{\rm ex}$ — is less well-constrained than in J1056+1419, and additional line detections, particularly at high excitation, would be required to confirm the temperature structure.


\subsection{H$_2$ Column Densities and Warm Molecular Gas Masses}
\label{subsec:h2props}
\begin{table*}
\centering
\caption{Column Densities and Molecular Gas Masses}
\label{tab:h2_colden_gasmass}
\begin{tabular}{lccccc}
\hline
\hline
Galaxy & $T_{\rm ex}$ (K) & H$_2$(1--0)S(3) & H$_2$(1--0)S(2) & H$_2$(1--0)S(1) & $M_{H_2(1-0)S(1)}$ ($M_\odot$) \\
 &  & $N_{\rm H_2}$ (cm$^{-2}$) & $N_{\rm H_2}$ (cm$^{-2}$) & $N_{\rm H_2}$ (cm$^{-2}$) & \\
\hline
J0754+3910 & 2000 & $<2.49\times10^{18}$ & $<5.85\times10^{18}$ & $<2.93\times10^{18}$ & $<2.79\times10^{4}$ \\
J0028+0055 & 2000 & $<3.60\times10^{18}$ & $<1.79\times10^{19}$ & $<8.31\times10^{18}$ & $<9.37\times10^{4}$ \\
J0809+3455 & 2000 & $<1.52\times10^{18}$ & $<9.28\times10^{18}$ & $<3.69\times10^{18}$ & $<2.51\times10^{4}$ \\
J1040+2957 & 4313 & $1.35\times10^{18}$  & $2.07\times10^{18}$  & $1.13\times10^{18}$ & $9.56\times10^{3}$ \\
J1056+1419 & 2063 & $2.14\times10^{19}$  & $1.58\times10^{19}$  & $1.25\times10^{19}$ & $1.26\times10^{4}$ \\
J0056$-$0936 & 2000 & $<4.99\times10^{17}$ & $<4.12\times10^{18}$ & $2.79\times10^{18}$ & $<3.08\times10^{4}$ \\
J0829+1754 & 2000 & $<2.37\times10^{18}$ & $<8.33\times10^{18}$ & $<3.07\times10^{18}$ & $<2.49\times10^{4}$ \\
J0758+2705 & 2000 & $<1.90\times10^{18}$ & $<9.09\times10^{18}$ & $<2.56\times10^{18}$ & $<2.60\times10^{4}$ \\
\hline
\end{tabular}
\tablecomments{
Column densities are computed assuming optically thin emission, an ortho-to-para ratio of 3, and a solid angle $\Omega = 0.13$ arcsec$^2$. Warm H$_2$ masses are derived from the H$_2$(1--0)S(1) line assuming local thermodynamic equilibrium (LTE). Excitation temperatures are measured for J1040+2957 and J1056+1419, and assumed to be $T_{\rm ex}=2000$ K for all other systems, following common practice for warm molecular gas studies \citep{mazz13, riff21}. Upper limits denote $3\sigma$ constraints. Einstein $A$ coefficients are taken from \citet{Wolniewicz1998}, and partition functions follow standard prescriptions.}
\end{table*}
Using the measured H$_2$ line fluxes and excitation temperatures derived in Sections \ref{subsec:emline} and \ref{subsec:h2excite}, we estimate the warm molecular gas column densities and masses under the assumption of optically thin ro-vibrational emission in local thermodynamic equilibrium (LTE). For a single excitation temperature $T_{\rm ex}$, the total H$_2$ column density is obtained from the population of the upper level $(v,j)$:
\begin{equation}
N_{\rm H_2} = N_{v,j} \frac{Z(T_{\rm ex})}{g_{v,j} \exp(-E_{v,j}/kT_{\rm ex})},
\end{equation}
\noindent where $Z(T_{\rm ex})$ is the partition function, $g_{v,j}$ is the statistical weight, and $E_{v,j}$ is the upper-level energy. The partition function is computed by summing over bound ro-vibrational levels assuming an ortho-to-para ratio of 3 appropriate for high-temperature equilibrium  \citep{wil2000}. The corresponding warm H$_2$ mass is
\begin{equation}
M_{H_2} = \frac{4\pi D_L^2 F_{ul}}{A_{ul} h\nu_{ul}} 
\frac{Z(T_{\rm ex})}{g_u \exp(-E_u/kT_{\rm ex})} m_{\rm H_2},
\end{equation}
\noindent where $D_L$ is the luminosity distance, $F_{ul}$ is the observed line flux, and $m_{\rm H_2} = 3.35 \times 10^{-27}$ kg is the molecular mass.

The inferred H$_2$ column densities span $N{_\mathrm{H_2}} \sim 10^{18}$--$10^{19}$~cm$^{-2}$, with warm H$_2$ masses in the range $M_{\mathrm{H_2}} \sim 10^{3}$--$10^{4}$~M$\odot$. These values are typical for the warm ($T \sim 2000$~K) molecular gas component traced by near-infrared ro-vibrational lines, and match those observed in the nuclear regions of nearby Seyfert and star-forming galaxies \citep{mazz13, riff21}.

\subsection{Shock Diagnostics}
\label{subsec:shocks}

\begin{figure*}
    \centering
    \includegraphics[width=\textwidth]{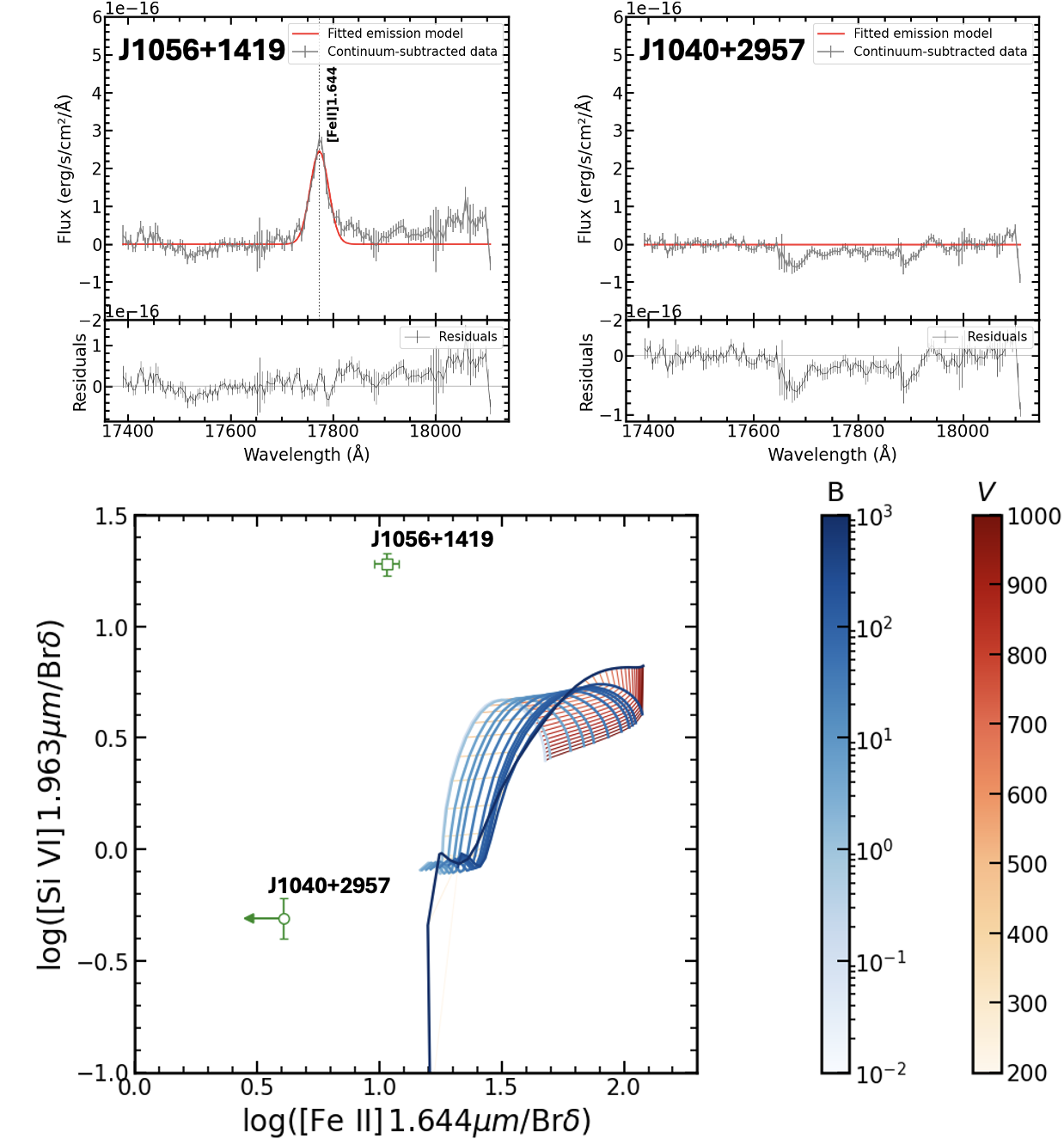}\\
     \caption{\textit{Top panels:} Continuum-subtracted GNIRS spectra around the [Fe\,II] 1.644\,\micron\ line for J1056+1419 \textit{(left)} and J1040+2957 \textit{(right)}, with best-fit emission models overplotted in red. J1056+1419 shows a clear detection, while J1040+2957 is dominated by noise with no significant line detection. Residuals are shown beneath each spectrum. \textit{Bottom panel:} Diagnostic diagram of $\log([\mathrm{Fe\,II}]\,1.644\,\micron/\mathrm{Br}\delta)$ versus $\log([\mathrm{Si\,VI}]\,1.963\,\micron/\mathrm{Br}\delta)$, overlaid with shock+precursor models from \citet{alle08} assuming solar abundances and preshock density $n = 10^3\,\mathrm{cm^{-3}}$. The model grid spans magnetic parameters $B/n^{1/2} = 10^{-4}$--$10\,\mu\mathrm{G\,cm^{3/2}}$ (blue) and shock velocities $v_s = 200$--$1000\,\mathrm{km\,s^{-1}}$ (orange/red). The green square marks J1056+1419, using the total integrated [Si\,VI] flux. The green circle marks J1040+2957. Due to the low signal-to-noise of the spectrum and the absence of a significant [Fe\,II] detection, we report a $3\sigma$ upper limit, shown as a leftward arrow.}
     \label{fig:shocks}
    
\end{figure*}


The H$_2$ excitation diagrams presented in \S\ref{subsec:h2excite} indicate that both systems host warm molecular gas with temperatures consistent with shock heating, and a clear multi-temperature structure characteristic of a complex shock environment. However, ro-vibrational H$_2$ emission alone does not uniquely constrain the dominant excitation mechanism, as both shocks and X-ray/UV irradiation from an AGN can produce elevated excitation temperatures \citep[e.g.,][]{maloney96, dwoods90} 
To disentangle these scenarios, we examine complementary diagnostics that probe distinct ionization regimes of the interstellar medium.

The combination of low-ionization lines such as [Fe\,II] and high-ionization coronal lines such as [Si\,VI] provides a powerful means of distinguishing between shock excitation and AGN photoionization. The [Fe\,II] emission traces partially ionized gas and is commonly enhanced in fast shocks due to grain destruction and collisional excitation \citep{alle08}, while [Si\,VI], with an ionization potential of 167 eV, requires a much harder radiation field and is typically associated with AGN photoionization or extreme shock conditions. As a result, line ratio diagnostics such as [Fe\,II]/Br$\delta$ and [Si\,VI]/Br$\delta$ provide a direct probe of the relative importance of shocks and photoionization.

To assess the origin of the ionized gas in our targets, we compare the observed line ratios to the shock+precursor models of \citet{alle08}, adopting solar abundances and a preshock density of $n = 10^3\,\mathrm{cm^{-3}}$. These models span a wide parameter space in magnetic field strength and shock velocity, allowing us to evaluate whether the observed emission can be reproduced by shocks alone.

Figure~\ref{fig:shocks} presents the resulting diagnostic diagram. For J1056+1419, the measured line ratios lie well above the model grid in $\log([\mathrm{Si\,VI}]/\mathrm{Br}\delta)$, indicating that the observed coronal-line emission cannot be reproduced by shocks within the explored parameter space. While shocks may still contribute to the excitation of the molecular gas and to the [Fe\,II] emission, the elevated [Si\,VI]/Br$\delta$ ratio requires a significantly harder ionizing spectrum, consistent with a dominant contribution from AGN photoionization. This interpretation is in agreement with previous studies of Mrk 273 \citep{Viv2013}, which find that such high coronal-line ratios are difficult to achieve through shocks alone and instead trace the presence of a luminous AGN.

In contrast, J1040+2957 remains unconstraining in this diagnostic space. We place a $3\sigma$ upper limit on the [Fe\,II] emission, and its location on the diagram is broadly consistent with the low-excitation end of the model grid within the uncertainties. As such, the available data do not allow us to distinguish between shock excitation and photoionization in this system. Deeper observations would be required to place meaningful constraints on its excitation mechanism. Taken together, these results suggest a picture in which shocks are likely present in the molecular gas phase, as indicated by the H$_2$ excitation, but are not sufficient to explain the full ionization structure of the gas in J1056+1419. Instead, the coronal-line emission points to a dominant AGN ionizing source, highlighting the coexistence of shock-heated molecular gas and AGN-driven photoionization within the same system.

As further supporting evidence, we also examine the ratio of ro-vibrational $\mathrm{H}_2\,(1\textrm{--}0)\,\mathrm{S}(1)$ to Br$\delta$ emission. Since both lines arise from gas at similar temperatures and are separated by only $\sim 50~\text{\AA}$ in the observed frame, this ratio is virtually free of differential dust extinction and probes the excitation conditions within the same gas clouds, avoiding the ambiguities associated with comparing lines from physically distinct ionization zones \citep{pux1990, left2024}. 

\citet{pux1990} show, for 44 H\,\textsc{ii} regions in 30 galaxies, that $\mathrm{H}_2\,(1\textrm{--}0)\,\mathrm{S}(1) / \mathrm{Br}\gamma$ ranges from $0.1$ to $1.5$ in star-forming regions (mean $0.5\pm0.3$), with values exceeding this range signifying shock-dominated excitation. After converting our Br$\delta$ fluxes to Br$\gamma$ using the Case B recombination ratio ($\mathrm{Br}\gamma / \mathrm{Br}\delta = 1.43$), we find $\mathrm{H}_2\,(1\textrm{--}0)\,\mathrm{S}(1) / \mathrm{Br}\gamma = 0.76 \pm 0.08$ for J1040+2957, and $7.0 \pm 1.3$ for J1056+1419. Notably, J1056+1419’s value is an order of magnitude above the star-forming range from \citet{pux1990}, matching values seen in confirmed jet-driven shock systems such as 3C 326 N \citep{left2024}, and providing strong, independent evidence of shock heating as already indicated by the $\mathrm{H}_2$ excitation diagram. Meanwhile, J1040+2957 falls within the higher end of the star-forming range; while shock contributions cannot be excluded, this ratio alone does not require them, consistent with the weaker and less well-constrained excitation structure in that system.

\section{Discussion}
\label{sec:discussion}

\subsection{Role of Mergers and Environment}
\label{subsec:mergers}

\begin{figure}
    \centering
    \includegraphics[width=0.45\textwidth]{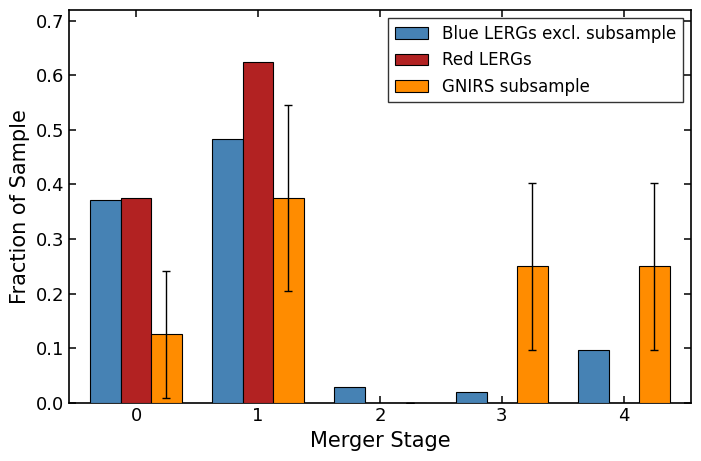}
    \caption{Fraction of galaxies as a function of merger stage for the parent blue LERG sample (blue), parent red LERG sample (red), and the BLERG subsample presented in this paper (orange). Stages are defined as: 0—undisturbed, 1—early interaction, 2—tidal features, 3—disrupted disks with two nuclei, 4—single coalesced nucleus. Error bars are $1\sigma$ binomial. Only the blue LERG samples are represented at the most advanced merger stages, with the BLERG subsample spanning both early and advanced stages but with larger uncertainty due to small sample size (see text for Kolmogorov–Smirnov test discussion).}
    \label{fig:merger_stages}
\end{figure}

\begin{figure*}
    \centering
    \includegraphics[width=\textwidth]{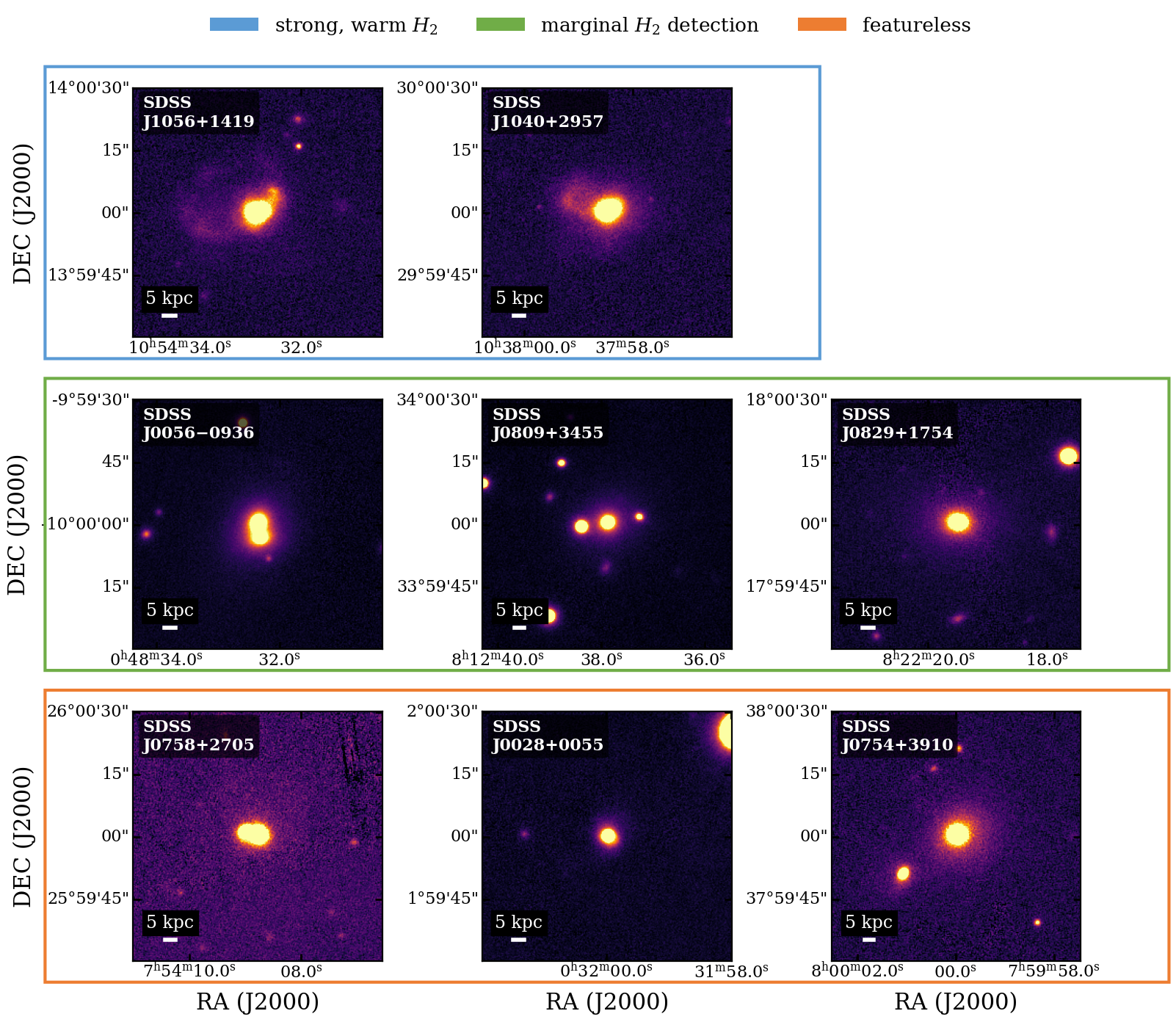}
    \caption{PanSTARRS $gri$ color composite images of the BLERG subsample discussed in this paper. Each panel shows one galaxy, with groups outlined according to their molecular hydrogen emission properties: strong warm H$_2$ (blue box), marginal H$_2$ detection (green box), and featureless (orange box). Each image is 1\arcsec\ $\times$ 1\arcsec\ with a scale bar of 5 kpc, and labeled with the SDSS designation. These images highlight the diversity of host morphologies as a function of H$_2$ excitation state.}
    \label{fig:panstarrs_h2}
\end{figure*}

The interplay between galaxy assembly via mergers, the triggering of star formation, and the growth of central supermassive black holes (SMBHs) has been extensively explored in theoretical frameworks \citep{Hopkins2006, dim2005} and remains an active area of observational study \citep{Ellison2011b}. Gravitational interactions are well-known drivers of enhanced molecular gas emission: systems such as the Taffy galaxies \citep{Zhu2007, Peterson2012} and Stephan's Quintet \citep{appl06, cluv2010} exhibit strong H$_2$ shocks leading to an excess of warm molecular hydrogen relative to other infrared cooling lines and PAH emission, suggesting that merger-driven turbulence can significantly alter the thermal state of the ISM independently of any AGN contribution. These systems provide a physical template for interpreting the ISM conditions of our BLERG sample, which we explored in detail in Sections~\ref{subsec:h2props} and \ref{subsec:h2excite}.

Here we ask whether the BLERGs in our sample are preferentially found in merging or interacting systems in order to understand whether and how the local environment modulates the ISM conditions of LERGs. We morphologically classify the full parent samples of blue LERGs and red LERGs using PanSTARRS \citep{PS1} $gri$ color imaging, following the five-stage merger classification scheme of \citet{petric2011, bridge2007}. Stages are defined as: (0) no clear evidence of disturbance; (1) early-stage interactions, where galaxies lie within 1 arcmin of each other but exhibit minimal or no morphological disruption; (2) systems displaying tidal features such as bridges and tails, while still lacking a shared envelope and retaining largely intact optical disks; (3) advanced mergers in which the original disks are disrupted yet two distinct nuclei remain identifiable; and (4) fully coalesced systems with a single merged nucleus.

Both blue and red LERGs from the parent sample show comparable fractions of systems in early or non-interacting stages (Stages~0--1; Figure~\ref{fig:merger_stages}). However, a key asymmetry emerges at later stages: only blue LERGs populate advanced merger stages (\(\geq2\)), comprising \(\approx 14\%\) of the parent blue LERG sample, whereas no such systems are identified among the red LERG population. This is consistent with the broader picture in which red LERGs are hosted by massive, dynamically relaxed, and isolated early-type galaxies whose gas reservoirs have long since been depleted \citep{best05, Heckman2014}, while blue LERGs appear to represent a population in which ongoing or recent interactions are delivering cold gas to the host, sustaining star formation and plausibly triggering the radio AGN episode \citep{jans2012, hardc2007}. 

The absence of late-stage mergers among red LERGs is particularly striking given that both populations span comparable stellar mass ranges \citep{best12}; it suggests that it is not mass alone but the presence of a dynamically disturbed, gas-rich environment that distinguishes blue from red LERGs. Figure~\ref{fig:panstarrs_h2} presents PanSTARRS gri color composites of the BLERG subsample, grouped by H$_2$ detection class (strong, marginal, featureless). These cutouts illustrate the full range of host morphologies and provide a visual complement to the merger stage distribution of our subsample.  

A two-sample Kolmogorov--Smirnov (KS) test confirms that the merger stage distributions of blue and red LERGs differ significantly (\(D=0.144\), \(p=0.009\)). For our eight-galaxy GNIRS subsample, the KS test yields \(D=0.384\), \(p=0.155\) relative to the parent blue LERG population, indicating no statistically significant difference, although the moderate KS statistic suggests some sensitivity to small-number statistics. In contrast, comparison with the red LERG population yields \(D=0.500\), \(p=0.026\), suggesting that the merger stage distribution of the GNIRS subsample differs significantly from that of red LERGs. Given the small size of the GNIRS sample, these results should be interpreted cautiously; nevertheless, they are broadly consistent with the subsample being more representative of the blue LERG parent population than of red LERGs. The morphological character of individual targets---in particular, the presence of tidal disturbance, close companions, or signatures of recent coalescence---therefore provides important context for interpreting the radio and molecular gas properties explored in the following sections.

\subsection{Radio Source Sizes and Population Statistics}
\label{subsec:jetsize}
\begin{figure}
    \centering
    \includegraphics[width=0.45\textwidth]{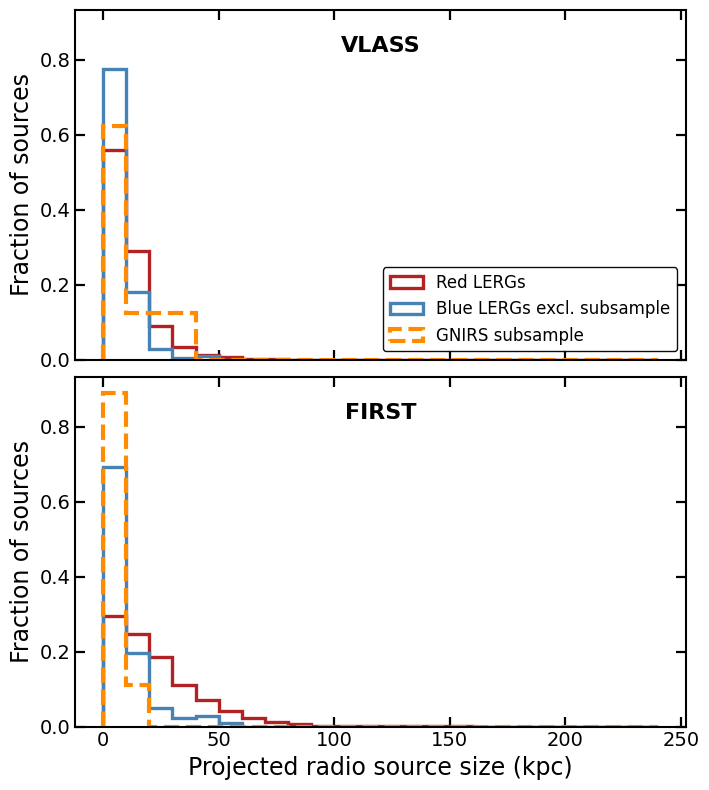}\\
     \caption{Distribution of projected radio-source physical sizes measured from the VLASS 2--4,GHz \citep{lacy2020} component catalog \textit{(top)} and from FIRST 1.4,GHz \citep{beck95}\textit{ (bottom)}. Red lines show the red LERG population, blue lines the blue LERG parent sample excluding our spectroscopic subsample, and the orange dashed line the BLERG subsample (N $=$ 8). Sizes are computed from the catalogs' deconvolved angular major-axis extents and converted to physical units using spectroscopic redshifts from the \citet{best12, crot06} parent sample. Histograms are normalized so the curves show the fraction of sources per bin.}
     \label{fig:jetsizeFIRSTVLASS}
\end{figure}

Radio-loud AGN are widely understood to inject mechanical energy into their surrounding environment through collimated jets---a process commonly referred to as kinetic or maintenance-mode feedback \citep[e.g.,][]{best12, Heckman2014, petric2003}. In this paradigm, radio jets inflate lobes that heat the surrounding hot gas halo, suppressing cooling flows and regulating star formation on halo scales. This mode of feedback is thought to be particularly dominant in LERGs, which accrete radiatively inefficiently and deposit most of their accretion energy as kinetic power into the surrounding medium \citep{hardc2007, best12}. A critical parameter governing its effectiveness is the physical scale reached by the jets: only jets whose extents are comparable to or exceed the host galaxy or its hot halo can couple efficiently to the circumgalactic medium and maintain quiescence in massive galaxies \citep{MCNN2007}. Physical size is therefore one of the most direct observational proxies for jet evolutionary stage and feedback capability.

Figure \ref{fig:jetsizeFIRSTVLASS} shows the distribution of radio source physical sizes for our eight-target BLERG subsample, together with the full blue LERG parent sample (171 targets, excluding our subsample) and the red LERG population (6,157 targets), all drawn from the VLASS 2--4\,GHz \citep{lacy2020} and FIRST 1.4\,GHz \citep{beck95} catalogs.

Physical sizes are computed from catalog deconvolved angular sizes and the spectroscopic redshifts from the \citet{best12} parent sample. The contrast between the populations is striking. Red LERGs span a broad, continuous distribution extending beyond 100\,kpc, consistent with mature, large-scale jets in quiescent elliptical hosts that have had sufficient time---likely several Gyr---to propagate well beyond the host galaxy \citep{best05, hardc2018}. The blue LERG parent sample is overwhelmingly concentrated in the first size bin ($\lesssim 20$\,kpc), with the distribution dropping by more than two orders of magnitude beyond 50\,kpc; only a modest tail extends to $\sim 50$\,kpc, indicating that large-scale jets are essentially absent in this population. A two-sample KS test applied to the VLASS blue and red parent samples confirms this quantitatively ($D = 0.301$, $p = 1.4 \times 10^{-16}$), strongly rejecting a common underlying size distribution.

Our eight-target BLERG subsample follows the same pattern. Five sources are morphologically unresolved, consistent with emission confined within the synthesized beam and physical sizes $\lesssim 5$\,kpc. The three spatially resolved sources---J0809+3455, J0829+1754, and J0056$-$0936---have deconvolved extents of 35, 26, and 19\,kpc respectively from direct SE image measurements, with J0056$-$0936 showing additional diffuse emission extending to $\sim 53$\,kpc in the median-stacked QL data (see \S\ref{subsec:subsamp_vlass}). Even adopting these larger direct measurements, all eight sources remain well below the 50--130\,kpc scales of the red LERG population. A KS test between the subsample and the red LERG population yields $D=0.4323$ with $p=0.0714$, and a KS test against the blue LERG parent sample gives $D=0.3216$ with $p=0.329$. Neither comparison is significant at the conventional $p<0.05$ level; taken together (and noting the small subsample size, $N=8$), these results are consistent with the BLERG subsample being drawn from the same underlying size distribution as the broader blue LERG population.

We repeat the size comparison using FIRST 1.4 GHz deconvolved sizes to test whether VLASS might be missing extended, lower-frequency emission. The FIRST results reinforce the VLASS conclusion: the blue LERG population is significantly more compact than red LERGs (KS test, $D=0.415$, $p=1.4\times10^{-33}$). Our BLERG subsample likewise differs from the red LERGs in FIRST ($D=0.696$, $p=7.1\times10^{-5}$) and shows only a marginal difference with the blue LERG parent sample ($D=0.457$, $p=0.037$). Thus the relatively small physical sizes we measure is not an artifact of VLASS's higher frequency/resolution---it persists in lower-frequency FIRST data that are more sensitive to extended, low-surface-brightness structure.

This small scale nature is not attributable to a difference in radio luminosity: both red and blue LERGs span overlapping ranges of $L_{1.4\,\mathrm{GHz}}$ in the \citet{best12} catalog, indicating that the size difference reflects a physical property of the host environment rather than an observational selection effect. Small scale radio extents of the jets is therefore a general characteristic of the star-forming LERG population, not a peculiarity of our eight targets.

\subsection{Subsample Individual Radio Morphologies}
\label{subsec:subsamp_vlass}
\begin{figure*}
    \centering
    \includegraphics[width=\textwidth]{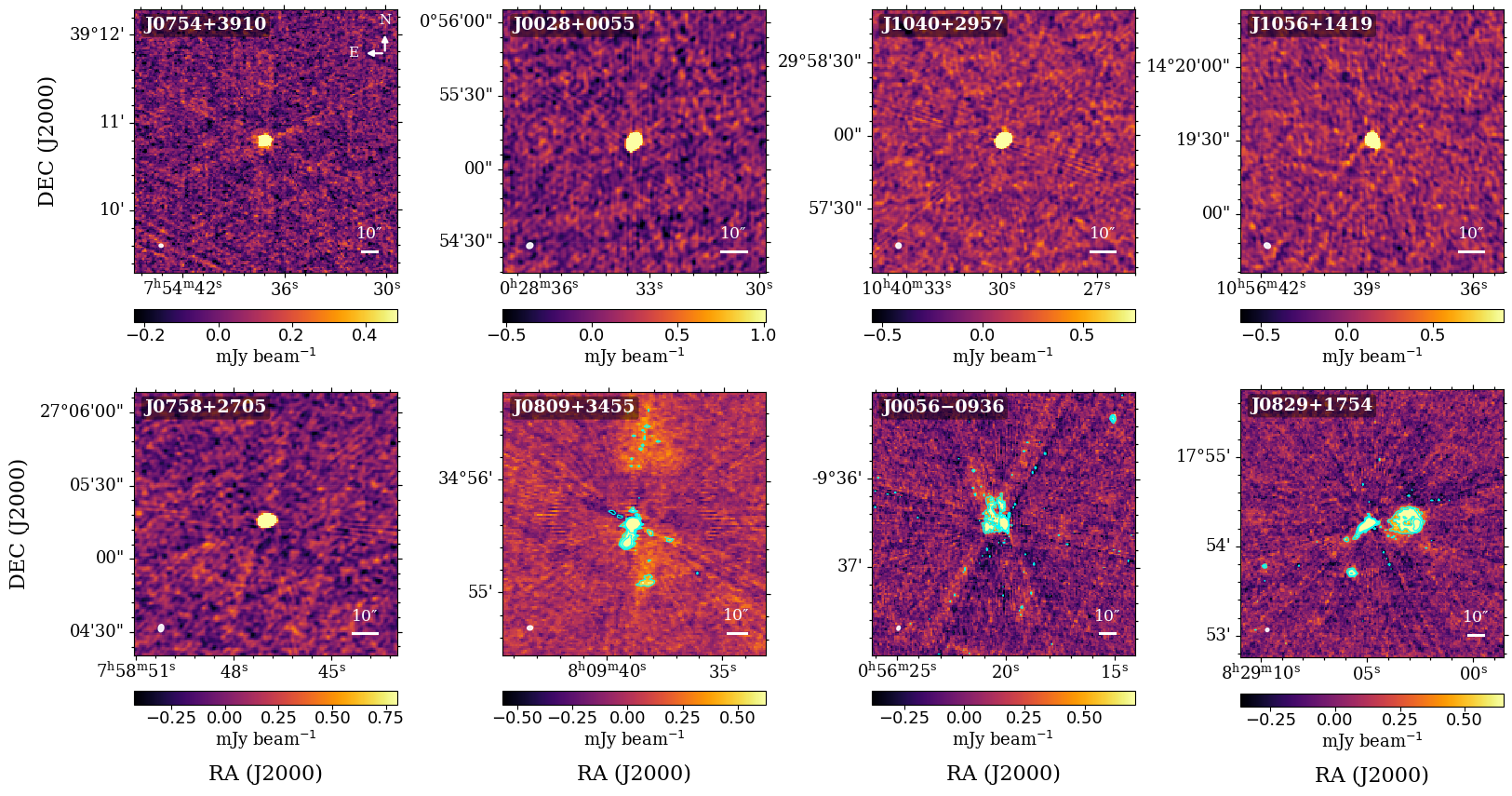}\\
     \caption{VLASS 2-4\,GHz continuum cutouts for the eight BLERG sources discussed in this paper, showing either median-stacked quicklook (QL) or single-epoch (SE) images. For point sources ---J0754+3910, J0028+0055, J1040+2957, J1056+1419, J0758+2705--- SE maps are used where available, capturing their small-scale, centrally peaked radio morphologies; J0754+3910 is shown in a QL coadd due to lack of SE imaging. For extended sources ---J0809+3455, J0056$-$0936, J0829+1754--- QL median images are used to enhance detection of faint, low-surface-brightness emission and recover the full spatial extent of resolved radio structures. QL coadds offer lower effective noise and improved sensitivity to diffuse emission compared to individual SE exposures, making them essential for characterizing jet and lobe morphology in these resolved cases. Cyan contours are overlaid for the extended sources at $3\sigma$, $5\sigma$, and $10\sigma$, where $\sigma$ is the local image RMS measured from an annular region in each panel. The color scale represents intensity in mJy beam$^{-1}$; the $10\arcsec$ scale bar and synthesized beam are shown in each panel. A compass indicating North and East is shown in the top-left panel and applies to all images.
}
     \label{fig:VLASScutouts}
    
\end{figure*}

Figure~\ref{fig:VLASScutouts} presents VLASS 2-4\,GHz continuum cutouts for all eight BLERG subsample sources. Five sources---J0754+3910, J0028+0055, J1040+2957, J1056+1419, and J0758+2705---are morphologically unresolved in all available single epoch (SE) and stacked quicklook (QL) images, with emission confined well within the synthesized beam, consistent with jets below the VLASS resolution limit of $\sim1$--2\,kpc at these redshifts. 

Relatively small radio extents does not, however, imply nuclear quiescence: J1056+1419 and J1040+2957 both show [Si\,\textsc{vi}]~1.9641\,$\mu$m detections in their $K$-band spectra (\S\ref{subsec:emline}--\ref{subsec:h2excite}), a 167\,eV ionization tracer attributable only to AGN photoionization or fast shocks; J1056+1419 additionally shows [Fe\,\textsc{ii}]~1.644\,$\mu$m and thermally excited H$_2$, indicative of shock-driven grain destruction and warm molecular gas at $T\sim2000$\,K in the nuclear ISM. These detections confirm that the AGN is already processing the nuclear gas in at least two small-scale jet sources, consistent with recently triggered radio AGN whose jets remain confined to the nucleus while the radiation field and nuclear-scale shocks begin to disturb the surrounding ISM. The remaining three sources exhibit resolved radio structure discussed individually below.

\paragraph{J0809+3455}
The VLASS image shows a clear bipolar morphology, with a bright central component and diffuse emission extending predominantly north-south, with an additional fainter extension to the west giving the source an overall cross-like appearance. The deconvolved extent is $23.3''$ (36~kpc) at $3\sigma$ and $18.8''$ (29~kpc) at $5\sigma$; the flux-weighted major axis of the bright inner structure is $\sim6.0''$ ($\sim$9~kpc). The bipolar emission is robustly confirmed at $5\sigma$, ruling out a noise artifact. The morphology — a bright core flanked by diffuse lobes — is reminiscent of a compact double or incipient FR\,I structure, consistent with jets propagating symmetrically into a roughly isotropic confining medium. At 36~kpc this is the most spatially extended source in the subsample, yet it remains well below the scales at which jets couple efficiently to the hot halo.

\paragraph{J0829+1754}
The VLASS image reveals an asymmetric structure with a bright core and a clear extension to the west-southwest, terminating in a secondary emission peak consistent with a hotspot or compact lobe terminus. The deconvolved maximum separation is $15.7''$ (26~kpc) at $3\sigma$ and $12.4''$ (21~kpc) at $5\sigma$; the QL coadd gives a consistent $\sim17.9''$ (30~kpc) at $3\sigma$. The flux-weighted major axis of $\sim4.2''$ ($\sim$7~kpc) indicates a compact but clearly resolved inner structure. The one-sided morphology and discernible working surface place J0829+1754 in the morphological class of compact symmetric objects (CSOs) or medium-symmetric objects (MSOs), interpreted as young radio AGN whose jets have driven compact lobes into the surrounding medium \citep{OdSa2021}. The asymmetric structure, rather than a symmetric double, is consistent with propagation into an inhomogeneous medium. At 26~kpc the jet has reached a scale comparable to the stellar half-mass radius of a typical massive galaxy at this redshift, but has not yet reached effective halo-coupling scales.

\paragraph{J0056$-$0936}
The SE image reveals a compact, centrally peaked source with modest resolved structure elongated approximately north-south, with a deconvolved maximum separation of $10.1''$ (19~kpc) at $3\sigma$ and $6.7''$ (13~kpc) at $5\sigma$. The flux-weighted major axis of $\sim2.0''$ ($\sim$4~kpc) at SE resolution indicates the inner structure is only marginally resolved, consistent with a compact or newly emerged radio source. The QL median coadd recovers additional low-surface-brightness emission extending to $28.5''$ (54~kpc) at $3\sigma$ and $22.8''$ (43~kpc) at $5\sigma$, suggesting extended diffuse emission on larger scales not captured in the SE image. The factor of $\sim3$ discrepancy in recovered extent between the SE and QL images is a sensitivity effect — the QL stack reaches a lower effective noise floor — rather than a calibration artifact. We caution that the QL images are not primary-beam corrected and stacking introduces correlated noise; confirmation of the full extent will require deeper single-epoch or targeted VLA observations. 

If confirmed, J0056$-$0936 would be the largest source in the subsample and the only target approaching scales at which jet--halo coupling becomes plausible, though 54~kpc remains at the low end of the range over which \citet{mcna07} argue jets become efficient feedback drivers. The $K$-band spectrum yields only a marginal H$_2$(1--0)S(1) detection with no higher-$J$ transitions (\S\ref{subsubsec:J0056}), providing no spectroscopic evidence for jet-driven shock enhancement at the current data depth. \\

Taken together, the three resolved sources span $\sim26$--53\,kpc while the five unresolved sources remain at $\lesssim5$\,kpc---a large internal spread within a homogeneously selected sample. This is consistent with the picture developed in \S\ref{subsec:mergers}: the blue LERG host environment is disproportionately associated with dynamically disturbed, gas-rich systems whose inhomogeneous ISM density distributions would cause jet propagation to vary stochastically between sources, with some jets breaking through to kiloparsec scales while others remain pinned at the nucleus. If jets in dense environments advance at $\sim0.01$--0.1$c$ \citep{OdSa2021}, the factor of $\sim5$--10 range in physical size implies a corresponding spread in effective jet age or confinement efficiency across the subsample.

The morphological diversity within the resolved sources bears on the question of whether these jets are genuinely young or environmentally frustrated. The symmetric bipolar structure of J0809+3455 is most naturally explained by propagation into a roughly isotropic medium. The one-sided hotspot of J0829+1754 and the asymmetric halo of J0056$-$0936 suggest instead asymmetric confinement or interaction with a clumpy circumnuclear ISM---precisely the conditions that hydrodynamic simulations predict will stall jet propagation and dissipate jet energy on sub-galactic scales \citep{bick2018, Mukherjee2018}. Both interpretations are physically plausible and likely operate simultaneously across the sample.

\subsection{Jet--ISM Coupling and Molecular Gas Excitation}
\label{subsec:radioH2}

\begin{figure*}
    \centering
    \includegraphics[width=\textwidth]{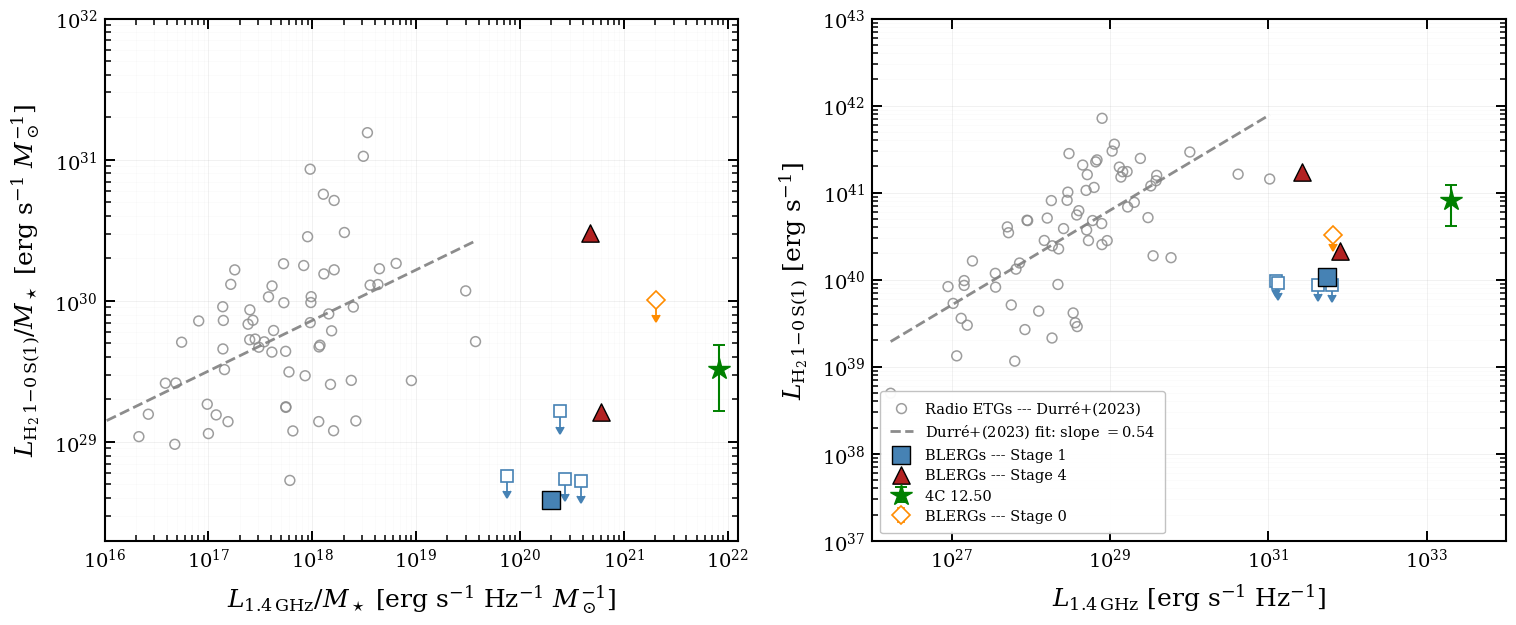}\\
    \caption{\textit{(left)} Mass-normalized and \textit{(right)} non-mass-normalized H$_2$ 1--0 S(1) luminosity versus 1.4\,GHz radio continuum luminosity for our BLERG subsample and comparison samples. All H$_2$ measurements trace the ro-vibrational 1--0 S(1) transition at 2.122\,$\mu$m, probing warm ($T \sim 2000$\,K) molecular gas. BLERGs (this work) are shown by merger stage: Stage~1 (blue squares), Stage~4 (red triangles), and Stage~0 (orange diamond). Filled symbols denote detections, and downward arrows indicate $3\sigma$ upper limits. Stellar masses are from MPA-JHU DR7 \citep{kauf03a, kauf03b, kauf03c}, and radio flux densities from NVSS \citep{Condon1998}. The comparison sample of radio early-type galaxies from \citet{Durre2023} (grey open circles; $N=68$ detections) is shown, with the dashed line indicating their best-fit relation in each panel. Also shown is 4C\,12.50 (green star), a radio-loud ULIRG at $z=0.122$ with confirmed jet--ISM interaction \citep{VillarMartin2023}. Its H$_2$ flux is from VLT/X-shooter spectroscopy, stellar mass from $K$-band photometry \citep{dasy11}, and 1.4\,GHz flux density from NVSS.}
     \label{fig:h2radio}
    
\end{figure*}

To investigate whether the small-scale jets in our BLERG subsample are already coupling to the molecular gas reservoir of their host galaxies, we examine the relationship between mass-normalized H$_2$ 1--0 S(1) luminosity ($L_{\mathrm{H}_2}/M_\star$) and mass-normalized 1.4\,GHz radio continuum luminosity ($L_{1.4\,\mathrm{GHz}}/M_\star$; Figure~\ref{fig:h2radio}). Mass normalization reduces spurious correlations driven by galaxy mass -- more massive galaxies tend to host both larger gas reservoirs and higher radio power -- but it also alters the interpretation: correlations in the normalized plane reflect changes in the quantity per unit stellar mass, whereas correlations in non-mass normalized space reflect changes in total luminosity. Because our BLERGs are interacting, star-forming systems that may not lie on the star-forming main sequence, we present both the mass-normalized (left panel) and the non-mass normalized (right panel) comparisons to ensure that normalization does not introduce or conceal trends. The mass normalized plot asks whether radio power and H$_2$ emission scale together per unit stellar mass, while the non-mass normalized plot asks whether galaxies with larger radio power simply host larger H$_2$ luminosities. 

All H$_2$ fluxes trace the ro-vibrational 1--0 S(1) transition at 2.122\,$\mu$m, probing warm ($T\sim2000$\,K) molecular gas and enabling a direct comparison across all datasets. Of our eight targets, three show detections of the H$_2$ 1--0 S(1) line: two Stage~4 merger systems (filled red triangles), detected in multiple H$_2$ transitions, and one Stage~1 close companion system (filled blue square). The remaining systems yield $3\sigma$ upper limits (downward arrows), including the Stage~0 system (open orange diamond). For comparison, we overlay the \citet{Durre2023} sample of nearby radio-active early-type galaxies (grey open circles), for which 68 systems show H$_2$ detections. 

Unlike studies based on \textit{Spitzer} mid-infrared rotational H$_2$ lines—which trace cooler gas and are more readily thermalized \citep{ogle10, ogle07, petric2018, hill14, lamb2019, minsley2020}—both our data and the \citet{Durre2023} sample use the same ro-vibrational 1--0 S(1) diagnostic, making this a direct comparison of the warm molecular phase. The \citet{Durre2023} sample spans a stellar-mass range of roughly $M_\star\approx4.9\times10^{9}$--$4.0\times10^{11},M_\odot$, so its upper mass range broadly overlaps the masses of our BLERG systems; however, the \citet{Durre2023} sample is dominated by quiescent early-type hosts, whereas our BLERGs are star-forming and interacting.


A key caveat in comparing total H$_2$ luminosities across samples is the strong dependence of H$_2$ surface brightness on the underlying excitation mechanism. As shown by studies of both star-forming and shocked systems (e.g., \citealt{herrera2012}), gas that is photoionized by AGN or star formation can produce much brighter H$_2$ emission than shock-excited gas for a given amount of molecular material. This implies that differences in $L{_\mathrm{H_2}}$ versus $L{_\mathrm{1.4 GHz}}$ between our BLERG sample and the \citet{Durre2023} radio early-type galaxy sample could reflect not only differences in the fraction of shocked gas but also in the relative importance of AGN or star-formation photoionization, unless this is explicitly accounted for. 


The \citet{Durre2023} sample exhibits a sub-linear correlation between $L_{1.4\,\mathrm{GHz}}/M_\star$ and $L_{\mathrm{H}_2}/M_\star$ (dashed line), indicating that galaxies with higher radio power emit relatively less warm H$_2$ per unit stellar mass. This behavior is consistent with their interpretation that radio activity and line-emitting phases are not tightly synchronized, producing only a weak time-averaged trend across the population. Our BLERG sample occupies the high-$L_{1.4\,\mathrm{GHz}}/M_\star$ end of this parameter space. However, their $L_{\mathrm{H}_2}/M_\star$ values lie at or below the extrapolated \citet{Durre2023} trend. In particular, the Stage~1 close companion system falls $\sim$1.5\,dex below the predicted relation, while the majority of the sample yields upper limits below the bulk of the \citet{Durre2023} distribution. Only the two Stage~4 merger systems approach the extrapolated relation from below. 

This is a notable result given the broader properties of our sample. BLERGs are actively star-forming systems undergoing mergers or close interactions, environments that are expected to host substantial molecular gas reservoirs. Nevertheless, they do not exhibit elevated $L_{\mathrm{H}_2}/M_\star$ relative to the radio ETG comparison sample. Instead, the data indicate that increased radio power per unit stellar mass does not correspond to a proportional increase in warm molecular gas emission. The internal variation within our sample further supports this interpretation. The two Stage~4 merger systems show higher $L_{\mathrm{H}_2}/M_\star$ than the Stage~1 system, suggesting that merger-driven processes—such as tidal shocks, gas inflows, and enhanced radiation fields \citep{guil09}—contribute to heating the molecular gas as interactions progress. This variation appears largely independent of the radio luminosity, which is comparable across the BLERG subsample.

The non–mass–normalized comparison in Figure~\ref{fig:h2radio} provides complementary context. In absolute luminosity space our BLERGs occupy H$_2$ luminosities that are at or below those of the \citet{Durre2023} radio–ETGs at comparable stellar mass, indicating that the low signal in the mass‑normalized plane is not simply an artifact of dividing by larger stellar masses in our sample. Instead, the deficit seen in $L_{\mathrm{H}_2}/M_\star$ largely reflects differences in host mass and/or differences in excitation efficiency (rather than a hidden absolute excess of warm molecular gas in the BLERGs). 


Additional context is provided by 4C\,12.50 (green star), a well-studied radio-loud ULIRG with confirmed jet–ISM interaction \citep{VillarMartin2023}. Despite having significantly higher $L_{1.4\,\mathrm{GHz}}/M_\star$ than our BLERGs, its $L_{\mathrm{H}_2}/M_\star$ is comparable to or lower than our detections and lies below much of the \citet{Durre2023} comparison cloud. As 4C\,12.50 is also a merging system, its H$_2$ emission likely reflects a combination of merger-driven and jet-driven processes. Its position on the diagram reinforces the conclusion that radio power alone does not set the level of warm molecular gas excitation.

Overall, while individual jets can couple strongly to the ISM on sub‑kpc–few‑kpc scales and may drive localized H$_2$ excitation in some objects \citep[e.g.,][]{Odea1998,holt08,morg13}, the ensemble behavior of our BLERG subsample favors a different conclusion: their warm H$_2$ emission is most consistently explained by merger‑ and interaction‑driven processes rather than by a dominant, population‑wide jet–ISM coupling. This interpretation is supported by (i) the lack of a positive internal correlation between radio power and H$_2$ (in both mass‑normalized and non-mass normalized space), (ii) the placement of BLERGs at or below the \citet{Durre2023} radio‑ETG locus, and (iii) the clear connection between H$_2$ strength and merger stage in our sample (see \S~\ref{subsec:mergers}). 

Nonetheless, given the prevalence of upper limits and the small sample size, individual objects (especially those with jets still confined within the dense ISM) remain candidate sites for significant jet–driven molecular excitation; spatially resolved follow‑up spectroscopy is required to distinguish these scenarios definitively. Furthermore, these processes need not be independent -- the same mergers that supply cold gas and power star formation can also trigger radio‑AGN activity, implying that merger‑ and jet‑driven heating may operate together in some galaxies \citep{best12,ell2015}.

\section{Summary}
\label{sec:conclusions}

In this paper, we present Gemini/GNIRS near-infrared spectroscopy of a sample of nine Blue Low-Excitation Radio Galaxies (BLERGs) --- star-forming radio AGN hosts drawn from the \citet{best2012, jans2012} parent catalog --- and combine these observations with morphological classifications and archival radio imaging to investigate the role of mergers, jet confinement, and ISM excitation in this rare population. Our results are summarized as follows:

\begin{itemize}
    \item We detect H$_2$ emission in five of eight targets: three showing detections in the 1--0 S(1), of which two show detections in multiple other H$_2$ transitions. 
    
    \item We morphologically classify the full parent samples of blue LERGs, red LERGs, as well as our subsample using PanSTARRS \emph{gri} color imaging following the five-stage merger classification scheme of \citet{bridge2007, petric2011}. Both populations show comparable fractions of systems in early or non-interacting stages (Stages 0--1). However, only blue LERGs populate advanced merger stages ($\geq 2$), comprising $\approx 14\%$ of the parent blue LERG sample, while no red LERGs are found at these stages. 

    \item We measure radio source physical sizes for the BLERG subsample and full blue and red LERG parent samples using the VLASS 2--4\,GHz and FIRST 1.4\,GHz catalogs. Red LERGs span a broad, continuous size distribution extending beyond 130\,kpc, consistent with mature large-scale jets in quiescent hosts. By contrast, blue LERGs --- including all eight targets in our subsample --- are overwhelmingly confined to $\lesssim 20$\,kpc, with the distribution dropping by more than two orders of magnitude beyond 50\,kpc. This small-scale nature is not attributable to a difference in radio luminosity, as both populations span overlapping ranges of $L_\mathrm{1.4\,GHz}$. We interpret the small-scale morphologies as evidence for either genuinely young radio AGN --- overlapping in size with Compact Steep Spectrum and Gigahertz-Peaked Source objects with kinematic ages of $10^3$--$10^5$\,yr --- or jets frustrated by the denser ISM characteristic of star-forming, merger-disturbed hosts, with both scenarios plausibly contributing simultaneously.

    \item We examine the relationship between mass-normalized H$_2$ 1--0 S(1) luminosity ($L_{\mathrm{H}_2}/M_\star$) and mass-normalized 1.4\,GHz radio luminosity ($L_{1.4\,\mathrm{GHz}}/M_\star$), finding that our BLERG sample occupies the high-radio-power end of the \citet{Durre2023} radio early-type galaxy comparison space, but does not show correspondingly enhanced $L_{\mathrm{H}_2}/M_\star$. Instead, the BLERG detections lie at or below the extrapolated \citet{Durre2023} relation, while several upper limits fall below the bulk of the comparison sample. Within our sample, the highest $L_{\mathrm{H}_2}/M_\star$ values occur in the most morphologically advanced merger systems, without a clear dependence on radio luminosity, pointing to merger-driven processes---such as tidal shocks, gas inflows, and evolving ISM conditions---as the dominant source of warm H$_2$ excitation.
    
    
\end{itemize}

Taken together, these results support the interpretation that BLERGs represent a brief but physically important transitional phase in galaxy evolution, in which a merger- or interaction-triggered radio AGN coexists with the gas-rich environment that spawned it. The same interaction that elevates ISM densities --- frustrating jet propagation and confining the radio source to sub-galactic scales --- also drives the H$_2$ through tidal shocks and merger-induced star formation. The rarity of BLERGs in the \citet{best12} catalog ($\approx 2.5\%$ of the LERG population) is consistent with this being a short-lived stage. As the jet matures and propagates beyond the dense circumnuclear medium, maintenance-mode feedback is expected to gradually quench star formation and move the host from the blue cloud or green valley onto the red sequence occupied by the vast majority of red LERGs. High-resolution radio imaging and spatially resolved molecular gas mapping of a larger BLERG sample would confirm whether individual systems sustain their AGN episodes long enough to complete this evolutionary transition.

\bibliography{SKbib, bibliography_petric, DBbio, sofia}{}
\bibliographystyle{aasjournal}

\begin{acknowledgments}

\end{acknowledgments}






\end{document}